\documentclass[acmtog]{acmart}
\acmSubmissionID{254}

\usepackage{booktabs} 

\usepackage[ruled]{algorithm2e} 

\SetAlFnt{\small}
\SetAlCapFnt{\small}
\SetAlCapNameFnt{\small}
\SetAlCapHSkip{0pt}

\setcopyright{acmcopyright}\acmJournal{TOG}
\acmYear{2020}\acmVolume{39}\acmNumber{6}\acmArticle{227}\acmMonth{12} \acmDOI{10.1145/3414685.3417816}

\usepackage{soul}
\usepackage{multirow}
\usepackage{tabularx}
\usepackage{graphicx}
\usepackage[export]{adjustbox}
\usepackage{hyperref}
\usepackage{amsmath}
\usepackage{graphicx}
\usepackage{tikz}
\usepackage{comment}
\usepackage{kbordermatrix}

\usepackage{adjustbox}

\newif\ifdraft
\drafttrue

\ifdraft
\newcommand{\nfc}[1]{{\color{cyan}\textbf{NF:} #1}}
\newcommand{\dcc}[1]{{\color{red}\textbf{DC:} #1}}
\newcommand{\rgc}[1]{{\color{green}\textbf{RG:} #1}}

\newcommand{\ahc}[1]{{\color{purple}\textbf{AH:} #1}}

\else
\newcommand{\nfc}[1]{}
\newcommand{\dcc}[1]{}
\newcommand{\rhc}[1]{}
\newcommand{\rgc}[1]{}
\newcommand{\sflc}[1]{}
\newcommand{\ahc}[1]{}

\fi

\usepackage[shortlabels]{enumitem}
                    \setlist[enumerate, 1]{1\textsuperscript{o}}

\begin{document}

\title{SketchPatch: Sketch Stylization via Seamless Patch-level Synthesis}

\author{Noa Fish}
\affiliation{
\institution{Tel Aviv University}
}
\authornote{Joint first authors}

\author{Lilach Perry}
\affiliation{
\institution{Tel Aviv University}
}
\authornotemark[1]

\author{Amit Bermano}
\affiliation{
\institution{Tel Aviv University}
}
\author{Daniel Cohen-Or}
\affiliation{
\institution{Tel Aviv University}
}



\begin{abstract}

The paradigm of image-to-image translation is leveraged for the benefit of sketch stylization via transfer of geometric textural details. Lacking the necessary volumes of data for standard training of translation systems, we advocate for operation at the patch level, where a handful of stylized sketches provide ample mining potential for patches featuring basic geometric primitives. 
Operating at the patch level necessitates special consideration of full sketch translation, as individual translation of patches with no regard to neighbors is likely to produce visible seams and artifacts at patch borders. 
Aligned pairs of styled and plain primitives are combined to form input hybrids containing styled elements around the border and plain elements within, and given as input to a seamless translation (ST) generator, whose output patches are expected to reconstruct the fully styled patch. An adversarial addition promotes generalization and robustness to diverse geometries at inference time, forming a simple and effective system for arbitrary sketch stylization, as demonstrated upon a variety of styles and sketches.
\end{abstract}

\begin{teaserfigure}
\newcommand{\vfig}{17}
\centering
\begin{tabular}[t]{c}
\includegraphics[width=\vfig cm]{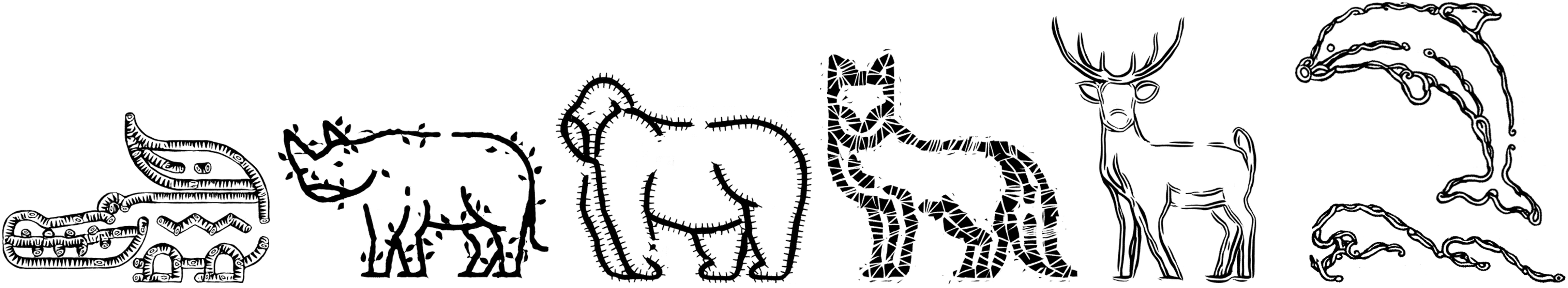} \\
\end{tabular}
\caption{A procession of stylized animal sketches generated by SketchPatch, from plain solid lined input sketches. Viewing on a monitor is recommended.}
\label{fig:teaser}
\end{teaserfigure}

\maketitle

\section{Introduction}
\label{sec:intro}

Visual creations generally carry elements of both content and style. Typically, the style elements are commonly linked to the rich combination of colors and textures, while the content ones manifest in the portrayed shapes. A sketch, on the other hand, differs in its stripped down and relatively simplistic construction, that is centered around the properties of its constituent lines and strokes. The thickness, angle, continuity and shape typically contribute more than texture to award the sketch its unique style, designed to appeal to the viewer. These differences are even more distinct when considering stylization. Unlike image stylization, which generally preserves the geometry and shape of the depicted scene, stylized sketches are often customized with the use of special brush strokes or repeating ornamental patterns.

These intricate patterns may be decorated with salient geometric details (see Figure
\ref{fig:teaser}), which are harder to capture and reproduce using image stylization techniques. Therefore, more often than not, decorative digital sketching relies on patterned brush tools. These are simplistic tools that allow the repeated stamping of predefined exemplars. These tools can be powerful and expressive, but require skill to master and are labour intensive. 
Attempts to automate this process have taken several routes. Some try to optimize the patterned stroke parameters \cite{kazi2012vignette}, and are hence forced to remain bound to the classic stroke model. Others explicitly define a content curve and a style one, and try to analyze the differences in curve behavior, which in turn are applied to any input content curve \cite{lang2015markov,lukavc2015brushables}. All of these approaches are limited in their ability to generalize, both in terms of difficult content curves, and in terms of involved styles. Hence, the most dominantly proposed direction is the exemplar-based one, where a library containing decorated or textured strokes is collected, and the strokes of a given sketch are fitted to their corresponding stylized ones from the library, optionally followed by stroke manipulation and refinement \cite{lu2013realbrush,lu2014decobrush}. These methods produce highly appealing decorated sketches, but still require manual specification of exemplar stroke directionality. 

In this paper, we propose a translation method that takes simple, unadorned lines and curves, and transforms them as indicated by a styled input exemplar. We take inspiration from neural style transfer methods \cite{gatys2016image,johnson2016perceptual,huang2017arbitrary,liao2017visual}. These methods allow the recombination, or transference, of stylistic details from a source exemplar onto the target, while simultaneously preserving its underlying content. These methods yield impressive artistic pieces, demonstrating competence on a wealth of image types and styles. However, since we aim for arbitrary sketch stylization, we must contend with highly varied geometries of strokes, and grapple with situations where both content and style are encoded in the geometry of the sketch, such that they are tightly entangled. This, together with data scarcity characterizing the type of intricate style elements we target, renders the usage of well-known image-to-image translation techniques intractable. 

Thus, instead of operating at the full image level, we advocate for operation at the patch level. We observe that, when broken down into pieces, sketch strokes disintegrate into a more tractable set of geometric primitives, which are more likely to be consistent across sketches of different geometries. Following this observation, we find that a rather limited set of consistently styled sketch exemplars (or even a single one), paired with their compatible plain counterparts, are often enough to provide a rich learning base for sketch stylization at the patch level. As such, we design a patch-to-patch translation model, that takes patches featuring plain geometric primitives and outputs their corresponding stylized version, while maintaining the underlying geometry. However, operating at the patch level necessarily raises the issue of noticeable seams at patch borders, when applied to a full sketch. 
To address this challenge, we propose generating context-aware patches, through the use of \textit{hybrid} patches. Hybrid patches are a combination of elements from a pair of plain and styled patches, where their central area stems from a plain element, and their boundaries could originate either from plain or styled patches (see Figure~\ref{fig:hybridization}). By training our model to generate corresponding fully styled patches, we are equipped to cater to arbitrarily sized sketch stylization via \textit{seamless translation} of overlapping sketch patches (see Figure \ref{fig:quilting}).

Given a full-sized unseen sketch at inference time, we propose circumventing the formation of contrasting patterns across different sketch regions through a \textit{pattern consistent} translation ordering. Patches comprising a full sketch are translated in a graph traversal order, rather than a raster order. This allows the generation of  consistent patterning, propagated along the graph, through the use overlapping patch generation (see Figure \ref{fig:test_time}).
Finally, to handle styled exemplars lacking paired plain counterparts, we also allow handling only roughly compatible plain exemplars. To do this, we rely on our patch-level geometric tractability observation and exploit the strong translation capabilities of a well established unpaired image-to-image translation model, namely CycleGAN \cite{zhu2017unpaired}. Through patch-level unpaired translation, we produce in preprocess, plain counterparts for any unpaired style exemplar, and thus are free to learn the translation between the original and generated sets.

We validate the performance of our solution for sketch stylization on an array of arbitrarily sized sketches, featuring an assortment of different geometries, and a variety of different styles.
We compare to two CycleGAN baselines and demonstrate their shortcomings with respect to geometry generalization and seamless translation. We further compare to classic translation \cite{hertzmann2001image}, and to optimization-based neural style transfer methods, illustrating that their reliance on image semantics is unsuitable for our problem setting. Our extensive experimentation demonstrates the robustness of our method to varying dataset sizes, different styles, and geometric sketch complexities, as well as the ability to generate seamless, content-aware stylization that is unique, aesthetic, and diverse. 

\begin{figure}[tb]
\newcommand{\plfig}{4.8}

\centering
\includegraphics[height=\plfig cm]{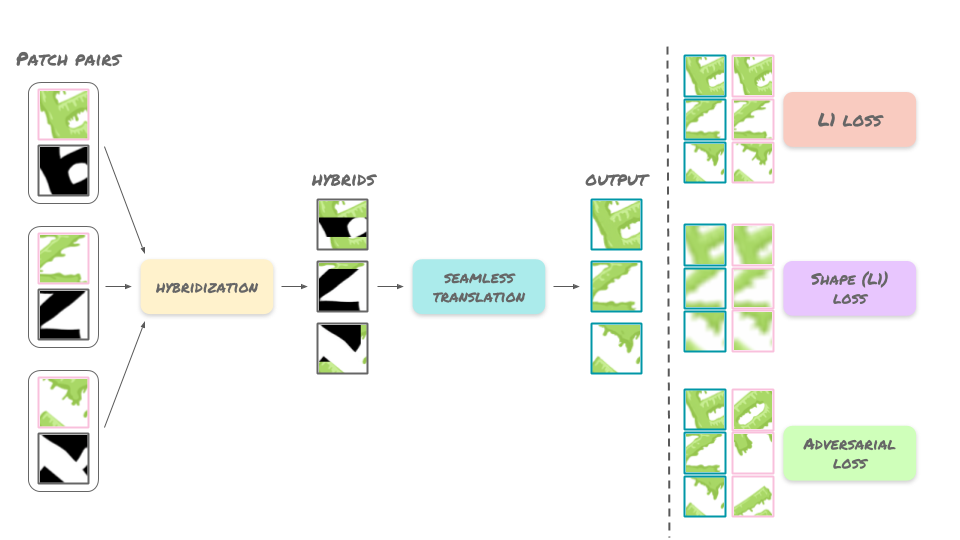}
\caption{Seamless translation pipeline. Pairs of styled (pink frames) and plain patches are combined to form hybrids that are given to the translation network, which outputs the corresponding fully styled patches (teal frames). A reconstruction (L1) loss compares the output to the original styled patch, while a shape reconstruction loss compares their blurred versions. An adversarial loss judges the output as well as randomly sampled real styled patches.}
\label{fig:pipeline}
\end{figure}
\section{Related Works}

Traditionally, synthesizing patterns along strokes from given exemplars includes three potential steps. First, a matching procedure selects the most fitting exemplar. Then, it is warped to match the given plain curve better, and finally, it undergoes further refinement.

Zhou et al. \shortcite{zhou2013example} break a given stroke into pieces, and then search for suitable exemplar parts to paste onto each piece while maintaining continuity along the interface between pieces. In RealBrush \cite{lu2013realbrush}, natural media samples are collected to form a library of brush strokes simulating real materials, facilitating digital painting. DecoBrush \cite{lu2014decobrush} builds upon the latter, and extends it to more intricate decorative patterns. It divides a user-defined sketch into segments, and matches each one to a patterned segment from a predefined stroke library. These are then warped to better reflect the original curvature of the input sketch, and are further refined using graph cuts and hierarchical texture synthesis.
Brushables \cite{lukavc2015brushables} takes as input an exemplar containing textural elements, and applies these elements to user paint strokes while preserving their global shape and structure. This method incorporates user interaction to determine the directionality of the textural elements.
Markov Pen \cite{lang2015markov} relies on a pair of curves, base and styled, defined by the user, and synthesizes stylized continuous strokes that follow a given target drawn path. 
Patternista \cite{10.5555/2981324.2981336} targets interactive ring-shaped object decoration that considers both style compatibility and spatial composition. 
It allows the user to select and place decorative elements upon a layout, and to synthesize fully decorated objects using a hidden Markov model.
gTangle \cite{santoni2016gtangle} procedurally generates tangle drawings using group grammars, allowing optional user control over tangle generation.
Wu et al. \shortcite{wu2018brush} take the neural approach to natural media stroke stylization, by first using physically based simulation for data generation, and then training a network to style a new stroke conditioned upon the current state of the canvas.

A survey on patch-based synthesis \cite{barnes2017survey} provides a more in-depth overview of relevant methods. 
All of these cases are either restricted in their expressiveness --- being able to use only existing patches, or rely on additional user data (e.g. additional curves or user-provided directionality) to drive the synthesis, which may be cumbersome to the novice user. Our approach, on the other hand, is fully automatic and profits from the creativity attributed to recent neural-based generation methods. 

\paragraph{Image-to-image translation}
Our solution to the described task can be cast as a sub-problem of the recently evolving field of image-to-image translation, aiming to transform images between two (or more) domains. An input image undergoes a process that simultaneously maintains certain visual properties, yet alters others, according to the source and target domains in question. Examples include grayscale to color, a semantic label map to an image, edge-map to photograph, etc. 
A pioneer in this field, Image analogies \cite{hertzmann2001image}, employs auto-regression to find patch-level analogies between images, such that the detected transformation can be applied onto a new image.  
Recently, however, these types of tasks are tackled almost exclusively using neural networks.

Typically, training such networks can be done in two manners, depending on data availability.
In the paired variety, a dataset containing pre-paired examples of images from the source and target domains, is given \cite{isola2017image}. Conversely, as the name suggests, in unpaired translation no such pairing is given, the two domains are presented separately, and a cycle consistency loss is introduced to regularize the transformation \cite{zhu2017unpaired,kim2017learning,yi2017dualgan,huang2018multimodal}. In our work, when pairing is unavailable, we use ideas inspired from traditional unpaired translation in order to generate pairs that can be used to train a paired-based system. As can be seen in Section \ref{sec:expr}, we demonstrate that this combination yields high-quality results. 

\paragraph{Neural style transfer}
Another sub-class of image-to-image translation that is similar to the question at hand is style transfer. 
In this paradigm, content and style elements are disentangled and recombined to generate differently styled versions of an input image, while maintaining its underlying content. 

Gatys et al. \shortcite{gatys2016image} utilize a pre-trained network (VGG \cite{simonyan2014very}) for feature extraction 
in an optimization process that fuses content from one image, with the style of another.
Deeper feature maps, known for capturing semantics, are used as content representatives, and shallow layers that capture stylistic elements, as style representatives (via the GRAM matrix).
Johnson et al. \shortcite{johnson2016perceptual} propose a feed-forward network setup and complement with a perceptual loss. 
Chen and Schmidt \shortcite{chen2016fast} enjoy the benefits of optimization allowing arbitrary style transfer, without compromising efficiency, by employing a patch-based swapping of style.
Huang and Belongie \shortcite{huang2017arbitrary} introduce a single network for arbitrary style transfer, providing
a simple yet highly effective component known as adaptive instance normalization, and Li et al. \shortcite{li2017universal} incorporate whitening and coloring transforms to modify content features to match the statistical characteristics of the style sample.
In Im2Pencil \cite{li2019im2pencil}, translation is used to re-style photos to appear hand-drawn, allowing user controllability for effects such as shading and sketchiness.  
Texler et al. \shortcite{texler2019enhancing} enhance the output of arbitrary style transfer methods via patch-based synthesis, addressing the lack of fine details that often characterizes their results.  
Finally, Liao et al. \shortcite{liao2017visual} utilize Patch Match \cite{barnes2010generalized} at different levels of the VGG feature pyramid, combined with an optimization process, to generate analogies of image pairs featuring semantically similar objects. 
Each image spawns an analogous image which resembles its immediate parent content-wise, but with style elements inspired by the other image.

The ideas proposed here are unfit for our task since they typically rely on rich images full of details and gradients, and are less proficient in changing the geometry of the content according to the input style. Similarly to the previous paragraph, we demonstrate this disadvantage in the context of our family of visual creations.

\paragraph{Text style transfer}
Our method is geared toward general sketch stylization and is not limited to text-based sketches. 
In our experiments, however, we make use of styled fonts to establish datasets of styled patches to learn from and transfer to arbitrary sketches at test time. 
Accordingly, we include a short overview of related methods that deal specifically with text style transfer.

Azadi et al. \shortcite{azadi2018multi} transfer the style of a few given glyphs, to new, unseen glyphs, by training a system composed of two networks that first predict glyph shape and then color and texture. The system learns the correlations between the 26 English letters and is therefore able to stylize any letter based on several given ones, which also limits the scope of the system to the specific letters (content) that were trained upon, in this case, English letters.
AGIS-Net \cite{Gao2019Artistic} and TET-GAN \cite{Yang2019TETGAN} also generate stylized glyphs from a small number of samples.
They disentangle the style and content of a glyph by leveraging an encoder-decoder architecture and recombining target glyph content with reference glyph style. These methods can be generalized to arbitrary writing systems. 
While this type of methods performs fantastically and can significantly aid even the most experienced typographer. These approaches are still limited to fonts, and hence are not appropriate to be adapted to our scheme. This is the case since these methods make critical use of the fact that the generated character is known and has a specific meaning that can be evaluated.

\section{Method}
\label{sec:method}

Geometric patterns, much like any other type of visual style, are abundantly available, yet individual styles do not normally boast many instantiations from which to learn. In classic textural style transfer, the essence of the style image can often be recovered from its statistics \cite{huang2017arbitrary}, thus a single image exhibiting a certain style may suffice, alleviating the demand for a substantial set of examples. 
In the realm of geometric elements within sketches, however, we are not privy to certain statistics or properties that capture the specific, and at times highly intricate, details of the style. Accordingly, as we aim to design a system that specializes in a certain geometric style for the purpose of sketch re-styling, we devise a method to mitigate the problem of data scarcity.

\subsection{Problem setting}
\label{sec:baseline}

We observe that, while most geometric styles are not instantiated sparingly, many of them are instantiated sufficiently for mining of basic primitives. That is, assuming a reasonably sized set of examples exists, 
one can cut a set of patches out of each example, and form a large collection of basic primitives that describe the given style. 

We treat the stylization problem as a translation problem, where one domain contains plain, unstyled stroke patches, and the other contains stylized stroke patches, as described by our given set of examples.
We assume the two domains are paired; each patch in the plain domain has a corresponding patch in the styled domain, and vice versa. 
In Subsection \ref{sec:unpaired} we describe our solution for style exemplars that are unpaired.

Unsurprisingly, the geometric diversity of the primitives within the training dataset directly influences the generalization capabilities of the translation model, and consequently, its performance at test time, under arbitrary sketches exhibiting arbitrary geometries. To that end, we seek to enrich the dataset further using augmentation.
Each example in our dataset is rotated in increments of $d$ degrees, and a patch is extracted every $p$ pixels (horizontally and vertically) on each ($d,p=8$ in our experiments).

Having assembled an appropriately sized and well-paired dataset, one can employ an image-to-image translation model such as Pix2Pix \cite{isola2017image}, to learn to transform plain strokes to styled ones. At inference time, any given sketch can then be cut into patches and translated patch-by-patch to form a styled sketch.
However, translating patches independently of their surrounding is likely to form visible seams at patch borders. While the appearance of these seams can often be reduced using post-processing stitching operations, we are still unable to guarantee a satisfactory result, for instance, when neighboring patches are stylized with misaligned or contrasting patterns (\textit{e.g.}, horizontal vs. vertical stripes).

\subsection{Seamless translation}

To handle border discontinuity and visible seams, we propose patch generation in an environment-aware manner. To do this, we train our translation network on the task of seamless translation, $ST$, that emphasizes and promotes smoothness and continuity of style elements, facilitating patch-by-patch translation of full-sized sketches.

\begin{figure}
\newcommand{\plfig}{6.1}

\centering
\includegraphics[width =\columnwidth] {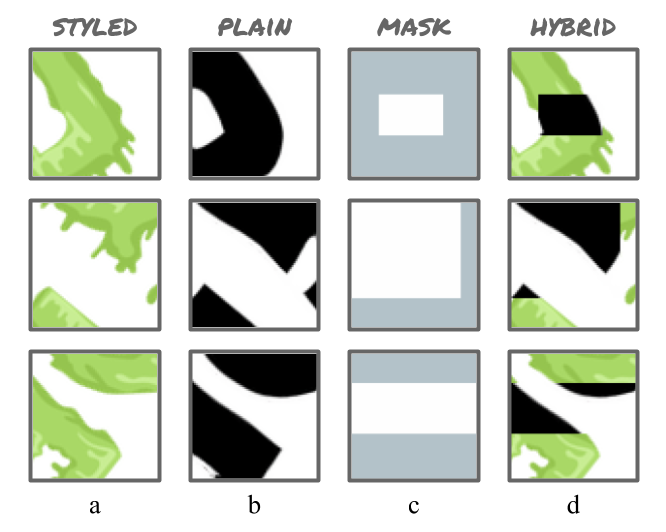}
\caption{Hybrid patch creation examples. In each row, a styled patch (a) and its corresponding plain version (b) are combined to form a hybrid (d), according to a hybridization mask (c), which is generated by randomly drawing four border integers, where gray indicates regions that are taken from (a), and white from (b).}
\label{fig:hybridization}
\end{figure}

We exploit the known pairing between patches in the plain and styled domains, and define a generator with an architecture that is identical to a single CycleGAN generator, which receives as input a hybrid patch, that is part styled and part plain, and outputs the corresponding fully styled patch (see Figure \ref{fig:pipeline}). To promote better generalization, the hybrid patch is generated on-the-fly, such that, at each iteration, we draw pairs of plain and styled patches 
and create a composition of the two in the following manner.
Our hybrid is initialized as the full plain patch.
We then randomly draw four integers $t, b, l, r$ in the range $[0,p/2-\delta]$, such that $p$ is the patch size and $\delta$ is a predefined threshold. We assign the value $0$ a $0.5$ probability of being selected, in order to ensure a reasonably sized plain region for the network to operate on, while the rest of the values are divided equally between the remaining $0.5$. 
These four integers determine the extent of the \textit{styled environment} at each of the four extremities of the patch: top, bottom, left and right.
With these in hand, we simply copy the corresponding sub-patch from the styled patch onto the hybrid. For instance, for a value $t$ indicating the top overlap, we take a sub-patch of size $t {\times} p$ from the top of the styled patch, and paste it onto the top of the hybrid.
This yields a hybrid that features plain sketch elements on the inside, and plain or styled elements at its extremities, depending on the overlap values that we have drawn (see Figure \ref{fig:hybridization}).

Our generator, $G$, is then trained to transform each hybrid into its fully styled version, which is of course known to us. Thus, $G$ employs a simple reconstruction loss via the $L1$ norm:

\begin{equation}
    \mathcal{L}_G = L_1(G(h),s),
\label{eq:recon}
\end{equation}

where $h$ and $s$ are the hybrid and styled patches respectively.

\begin{figure}
\newcommand{\ttrfig}{2.4}
\newcommand{\ttbfig}{3}
\setlength\tabcolsep{1pt}

\centering

\begin{tabular}[t]{c c c}

\includegraphics[height=\ttrfig cm]{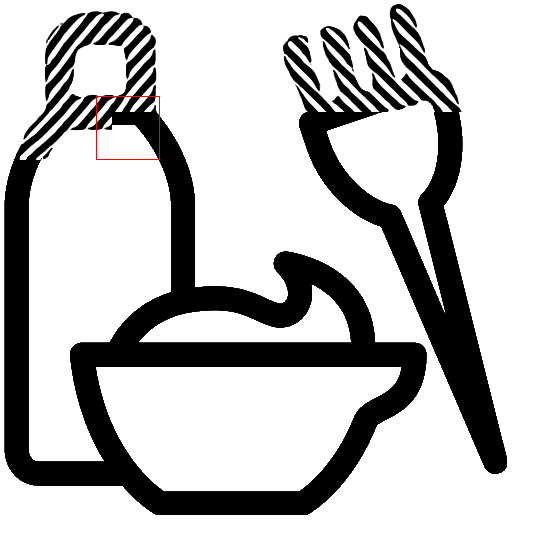} \hspace{8pt} &
\includegraphics[height=\ttrfig cm]{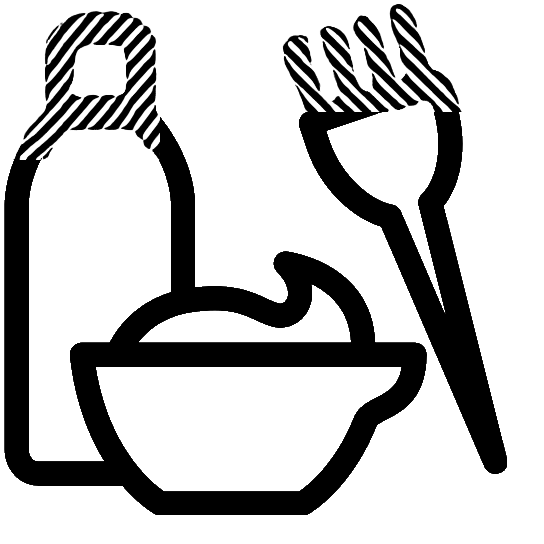} \hspace{8pt} &
\includegraphics[height=\ttrfig cm]{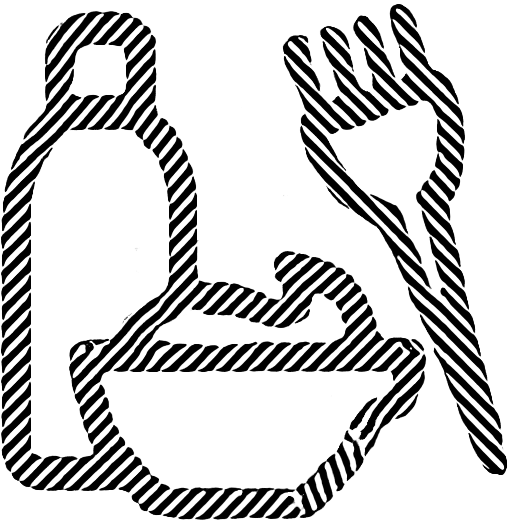} \\ 
a \hspace{8pt} & b \vspace{8pt} & c \vspace{8pt} \\

\end{tabular}

\begin{tabular}[t]{c c}
\includegraphics[height=\ttbfig cm]{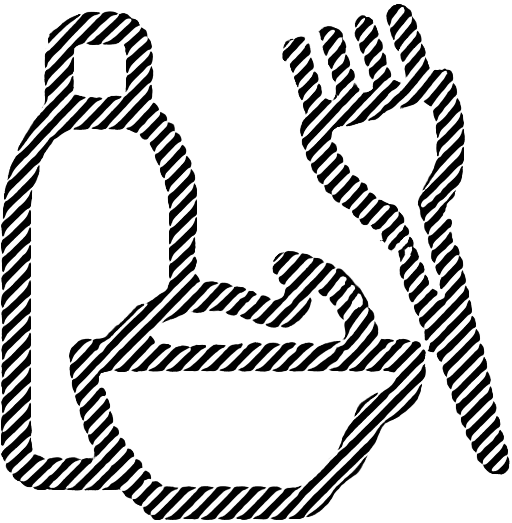} \hspace{20pt} &
\includegraphics[height=\ttbfig cm]{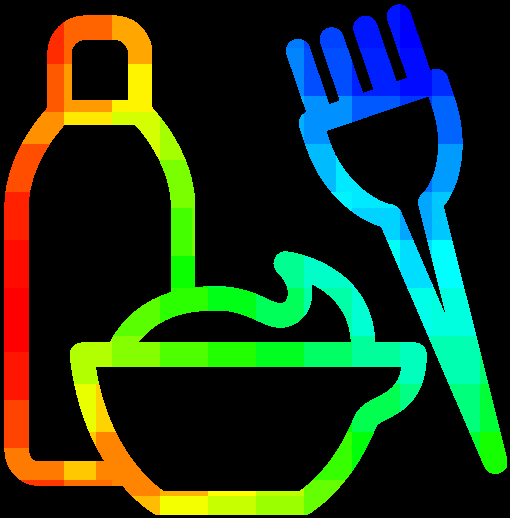} \\ 
d \hspace{20pt} & e \\

\end{tabular}
\caption{Test time translation. Red square in (a) marks the next patch to be translated, featuring styled elements at its top and left borders. The result is shown in (b). Full raster order translation result is shown in (c), exhibiting inconsistent stripe orientation as a result of sketch discontinuities leading to multiple independent translations. A consistent result is shown in (d), where graph traversal order ensured a dependent translation. The traversal order is visualized by a gradient color change from blue to red in (e). \textit{Icon made by \href{https://www.flaticon.com/authors/freepik}{Freepik}.}}
\label{fig:test_time}
\end{figure}

\begin{figure}
\newcommand{\plfig}{4.75}

\centering
\includegraphics[height=\plfig cm]{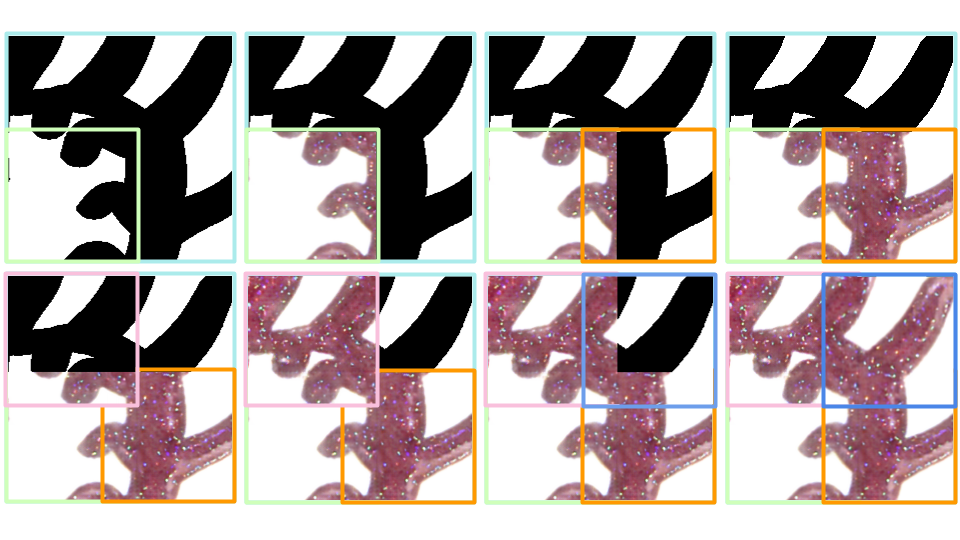}

\caption{Overlapping translation process. A 2x2 patch region undergoes translation starting at the top left, where the first patch to be translated is marked in green, and its translation result is shown right beside. In the third image (top row), the neighboring patch, marked in orange, is the next to be  translated. Sharing an overlap with its neighbor on the left, this patch goes in as a hybrid, and its translation is shown next. The process proceeds in the second row in a similar fashion.}

\label{fig:quilting}
\end{figure}

\subsection{Inference time}
At inference time, any given sketch is first cut into overlapping patches. That is, for patch size $p$ and overlap extent $o$ ($p=64, o=16$ in our experiments), we define a patch at each coordinate $((p-o) \cdot i, (p-o) \cdot j)$ for $i=0,..,\lceil\frac{h}{p-o}\rceil-1$ and $j=0,..,\lceil\frac{w}{p-o}\rceil-1$, where $h$ and $w$ are the height and width of the sketch. Undersized patches at the margins are padded with the background color of the sketch.

To translate these overlapping patches and generate the final styled sketch, we can simply follow a raster order and translate each patch as a hybrid of itself, with the already styled regions that it shares with the neighbors that precede it in the raster order (see Figure \ref{fig:test_time}(a-b)).

Following a simple raster order, however, may prove detrimental for certain styles and sketches. For instance, given a sketch with a discontinuous geometry in the raster sense, such that multiple patches that belong to the same connected component are translated "blindly" (with no styled portions for guidance), and given a style featuring continuous elements with a specific orientation, raster order translation will inevitably lead to independent creation of style elements at different sections of the sketch. These may clash with one another and ultimately compromise the quality of the outcome (see Figure \ref{fig:test_time}(c)).

To overcome this, and minimize the effects of independent translation, we compute a graph traversal order, and follow it to promote a more reliable and balanced translation. Specifically, we define a graph $G = (V,E)$, such that $V$ includes all the non-empty sketch patches, and $E$ contains an edge for any $i,j \in V$ whose corresponding patches are neighbors in the sketch, with a non-empty overlap region. We then select a root patch at random, and compute a traversal order starting at the root, using \textit{Breadth First Search} (BFS). In this manner, any connected component within the sketch will have a single independent translation -- that of the root, and any patch thereafter will be translated as a hybrid made up of itself and the overlapping styled regions of the neighbors that precede it (see Figure \ref{fig:test_time}(d-e)).
Note that as we train on rotation-augmented data, this approach does not guarantee preservation of the original \textit{global} orientation of the pattern.

Figure \ref{fig:quilting} demonstrates the overlapping translation process up close on a 2x2 patch region, starting at the top-left, and proceeding in a row-major order.

\subsection{Adversarial setup}
\label{sec:disc}

Having augmented our dataset in Subsection \ref{sec:baseline} for generalization purposes, we note that our network is put under increased strain, such that it struggles to learn guided solely by our original reconstruction loss (see Equation \ref{eq:recon}). This loss demands a full reconstruction of the styled patch, which may be unnecessarily constricting. Indeed, we wish to preserve the already styled regions of an input patch and seamlessly extend the translation into the plain regions, but we do not require an exact preservation of the original patch, as long as the plain regions are translated with faithfulness to the presiding style.

To address the increased training strain and loosen the overly strict reconstruction demands, we add an adversary -- a discriminator ($D$), to our ST system. Architecturally, $D$ is identical to a single CycleGAN discriminator (PatchGAN), and it employs a standard LSGAN \cite{mao2017least} discriminator loss, with the output of $G$ taken as \textit{fake}, and real patches from the pool of fully styled patches taken as \textit{real}. We update the loss of $G$ from Equation \ref{eq:recon}, to include the appropriate adversarial loss as well:
 
\begin{equation}
    \mathcal{L}_G = L_1(G(h),s) + (D(G(h)) - 1)^2
\label{eq:g_loss}
\end{equation}

The addition of $D$ assists in training $G$ in a more robust manner, decreasing the emphasis on strict reconstruction and allowing the network to generalize better to different geometries. 
However, as it decreases the weight put upon reconstruction, we find that $G$ is under increased risk of degeneration. That is, patches in the data contain expanses of varying magnitude of empty space, therefore $D$, naturally, recognizes empty space as \textit{real}. This, in turn, may cause $G$ to resort to taking the "easy" way out by synthesizing empty space in place of more intricate texture. To that end, we introduce an additional loss to help preserve the general shape of the primitive within the patch, without resorting to increasing, yet again, the strictness of the reconstruction loss by weighting it higher. We achieve that by introducing a \textit{shape} reconstruction loss, that passes the output patch and its corresponding fully styled patch through a Gaussian filter, and compares the two, thereby demanding reconstruction of the general shape of the stroke, rather than its exact minute details. The total loss of $G$, which extends the one described by Equation \ref{eq:g_loss}, is therefore:

\begin{equation}
    \mathcal{L}_G = L_1(G(h),s) + (D(G(h)) - 1)^2 + L_1(g(G(h)),g(s)),
\label{eq:g_loss_blur}
\end{equation}

where $g$ is a Gaussian filter with a kernel of size 10 and $\sigma=10$.

\subsection{Unpaired translation}
\label{sec:unpaired}

Bespoke style exemplars can easily be paired with their plain counterparts during the creation process. Conversely, in most cases, ready-made exemplars that can be found online, for instance, lack such a pairing. Many of them feature original and engaging patterns that could provide ample data for stylization, thus we choose to address their unpaired status here.

One option is to manually generate the plain counterpart by using a drawing application and a plain brush to re-trace the lines and curves of the underlying sketch within the exemplar. This is not a particularly tedious process, yet it still requires the right software, some minimal drawing skills, and perhaps even designated hardware, such as a drawing pen or a graphics tablet.

Therefore, we propose a second option, based on the unpaired image-to-image translation paradigm, specifically, CycleGAN \cite{zhu2017unpaired}.
In this scenario, we have on hand a styled exemplar (or a small set of consistent exemplars), which we have cut up into patches as before, but we are in need of a plain domain, containing similar patches to those in our styled domain, only featuring plain, unadorned strokes. Fortunately, sketches that are composed of plain strokes are prevalent, allowing us to assemble a collection of plain basic primitives by cutting up patches from a set of sketches that we either collect manually (similar to the target domain, it need not be large), or automatically, assuming possession of prior knowledge about our pool of sketches, such as its underlying geometries and stroke weight.

\begin{figure}[tb]
\newcommand{\plfig}{4.2}

\centering
\includegraphics[width=\columnwidth]{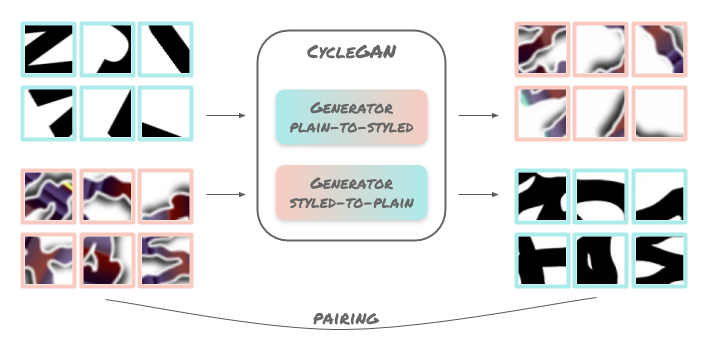}
\caption{Pairing unpaired patches. Unmatched plain and styled patches are given to CycleGAN to learn the translation between them. The trained styled-to-plain generator is used to obtain the plain version of all styled patches, thereby providing the pairing needed for training our ST network, when such is unavailable.}
\label{fig:cyclegan}
\end{figure}

Using the two separate yet geometrically similar domains, we train CycleGAN to learn the translation between plain and styled primitives. The trained model is then used to generate a well-aligned plain patch counterpart for each styled patch (by forwarding each styled patch through the styled-to-plain generator), thereby forming a paired dataset upon which our $ST$ model can be trained as before. See Figure \ref{fig:cyclegan}.

\section{Experiments}
\label{sec:expr}

We collected a diverse set of test sketches featuring varied geometries, and a set of geometric styles with which to examine the performance of our proposed solution.
The styles we experimented with come from different sources, some are paired with their corresponding plain counterparts, and some are not. Those that are, were trained only using the paired portion of our pipeline, namely, the ST network. The unpaired ones were roughly matched with plain exemplars featuring strokes of similar geometries, and were first passed through the patch-based CycleGAN discussed in Subsection \ref{sec:unpaired} in order to generate aligned pairs. These pairs were then passed through the rest of the pipeline (ST).

Our paired examples include styles designed by an artist and a novice user, and our unpaired examples include decorative typefaces that we collected online, as well as novice user creations and some miscellaneous sources.

Each exemplar (or set of consistently styled exemplars), was cut into patches, using rotations for augmentation, such that the pool of patches totaled at roughly $\sim$150k-300k patches. Paired examples were cut consistently (styled and plain) to preserve the alignment.

Since the collected styled typefaces are unpaired, each was assigned a plain typeface that is geometrically similar to it, following a brief visual comparison. We found stroke weight to be particularly important when matching styled and plain domains, due to CycleGAN's general shape preservation tendencies. 
Unpaired user-created exemplars were matched with a stroke library featuring similar geometries. 

We begin by performing a qualitative ablation experiment in Subsection \ref{sec:abl}. 
In Subsection \ref{sec:comp}, we present a variety of comparisons to our baseline approaches, as well as to neural style transfer methods, and in Subsection \ref{sec:results}, we provide further results obtained with our pipeline, offering insights into its strengths and weaknesses.

Note that all results are best viewed on a screen, while zooming-in to observe the details. 
Please refer to our supplementary material for comparisons  to Pix2Pix \cite{isola2017image} and to two font stylization techniques (\cite{azadi2018multi,Gao2019Artistic}), and for further results, as well as a short video clip.
Our code and data can be found on our Github page.

\subsection{Ablation}
\label{sec:abl}
To evaluate the validity of our full proposed solution presented in Section \ref{sec:method}, we trained its partial versions and examined their respective performances. The first is the most basic, featuring our generator $G$ trained using only a simple reconstruction loss. The second version adds a discriminator and its corresponding losses. 
The third incorporates both standard reconstruction and shape reconstruction (discussed in Subsection \ref{sec:disc}, and the final one features our full solution, including all three losses.

\begin{figure}
\newcommand{\ttrfig}{3.7}
\newcommand{\ttsfig}{1.2}
\setlength\tabcolsep{1pt}

\centering
\begin{tabular}[t]{c c c c c}

\includegraphics[height=\ttsfig cm]{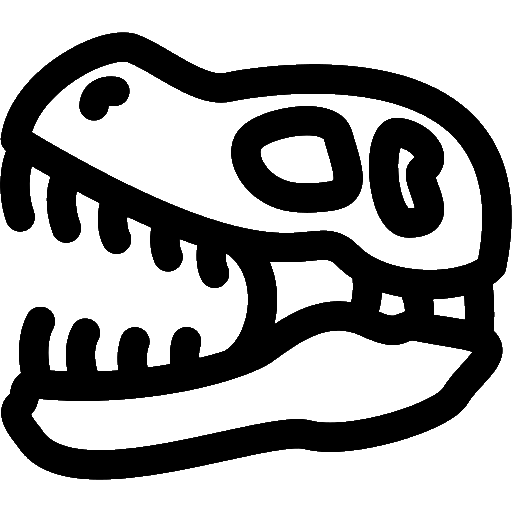} &
\includegraphics[height=\ttsfig cm]{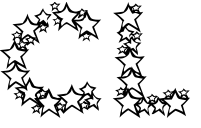} &
\hspace{8pt} &
\includegraphics[height=\ttsfig cm]{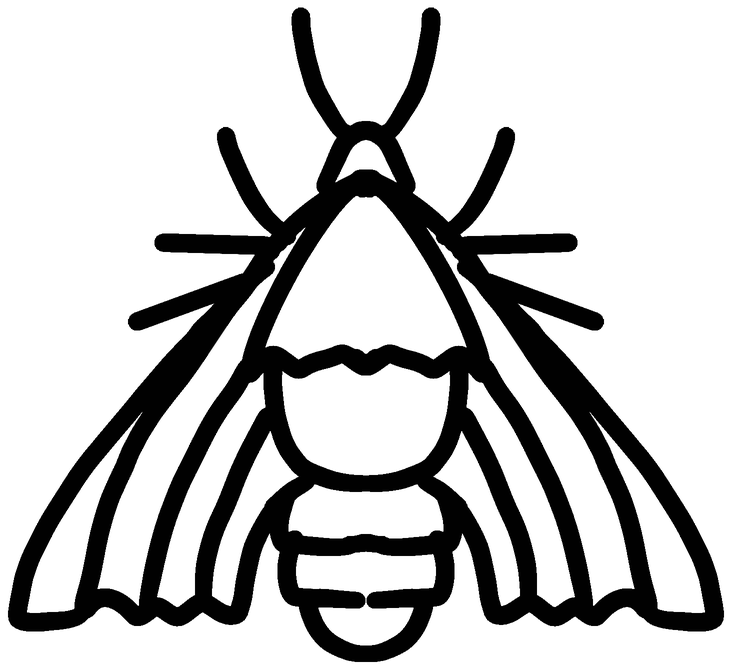} &
\includegraphics[height=\ttsfig cm]{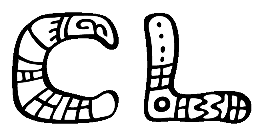} \\
\hline
\\
\multicolumn{2}{c}
{\includegraphics[height=\ttrfig cm]{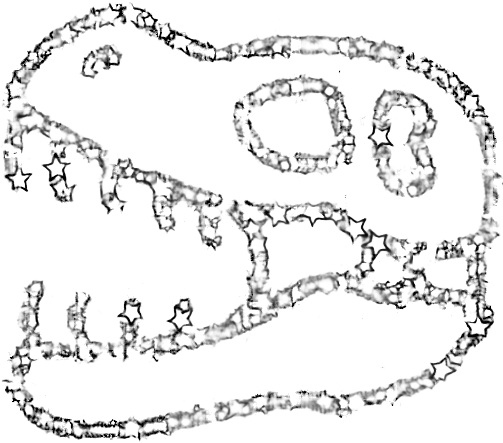}} &
\hspace{8pt}  &
\multicolumn{2}{c}
{\includegraphics[height=\ttrfig cm]{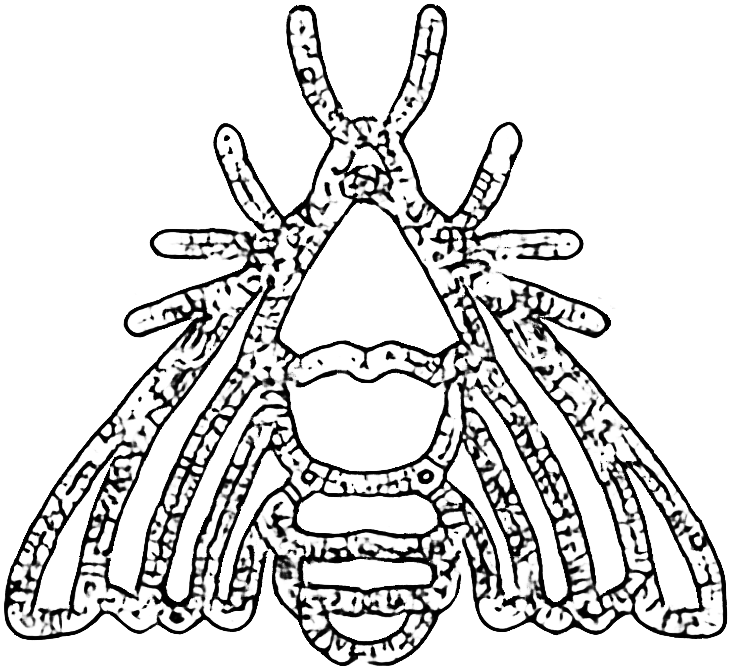}} \\
\multicolumn{2}{c}{(a)} &
\hspace{8pt} &
\multicolumn{2}{c}{(b)} \\
\multicolumn{2}{c}
{\includegraphics[height=\ttrfig cm]{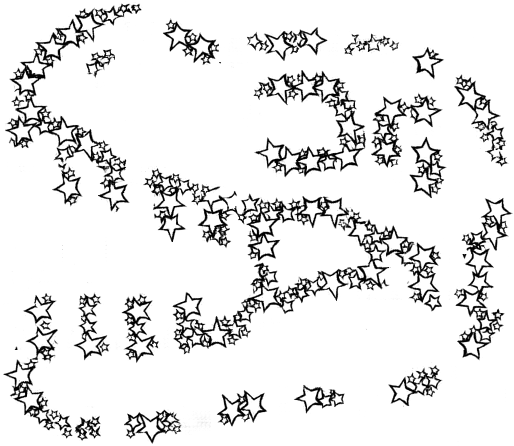}} &
\hspace{8pt} &
\multicolumn{2}{c}
{\includegraphics[height=\ttrfig cm]{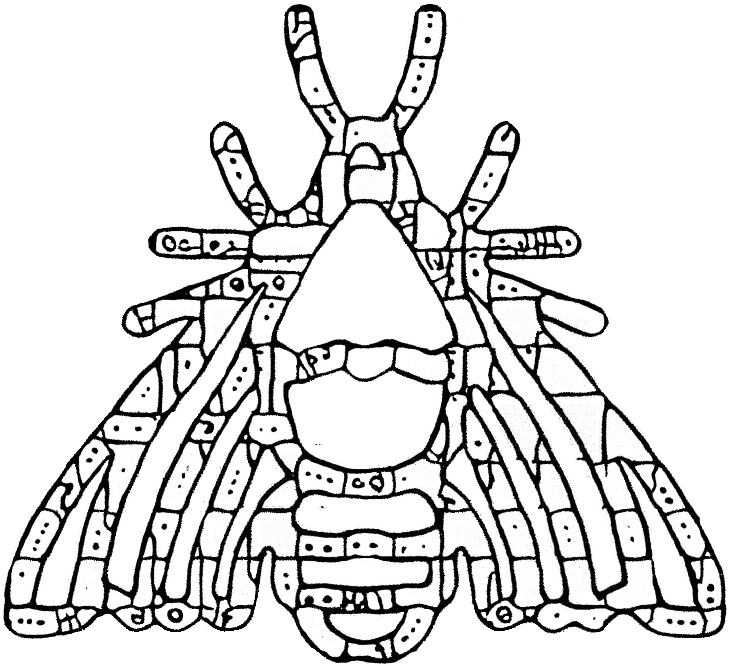}} \\
\multicolumn{2}{c}{(c)} &
\hspace{8pt} &
\multicolumn{2}{c}{(d)} \\
\multicolumn{2}{c}
{\includegraphics[height=\ttrfig cm]{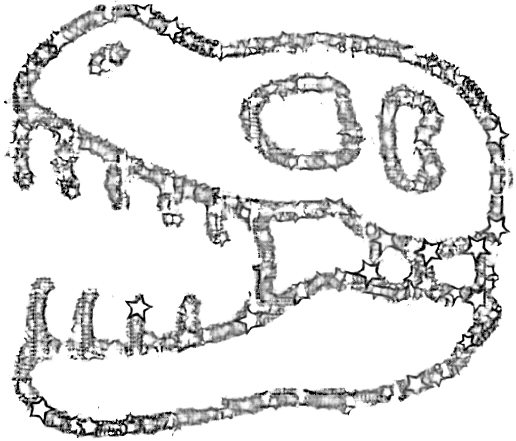}} &
\hspace{8pt} &
\multicolumn{2}{c}
{\includegraphics[height=\ttrfig cm]{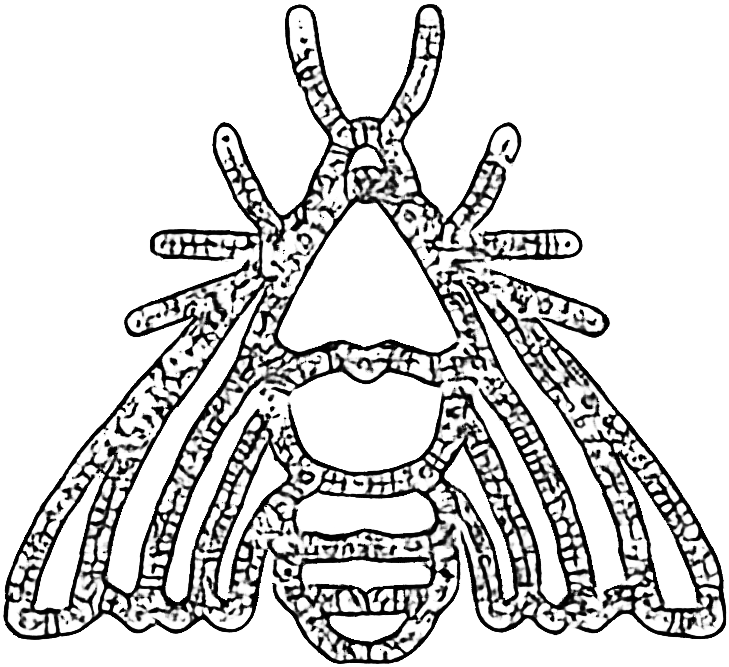}} \\
\multicolumn{2}{c}{(e)} &
\hspace{8pt} &
\multicolumn{2}{c}{(f)} \\
\multicolumn{2}{c}
{\includegraphics[height=\ttrfig cm]{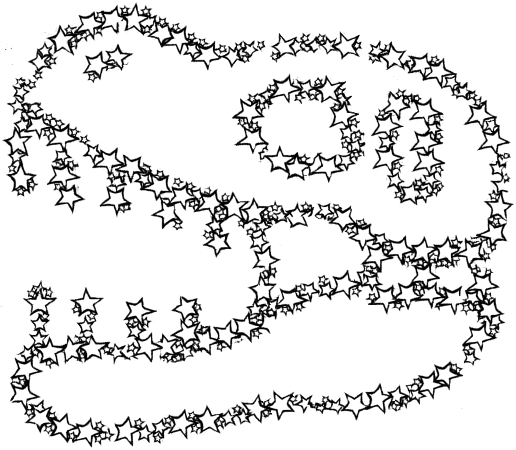}} &
\hspace{8pt} &
\multicolumn{2}{c}
{\includegraphics[height=\ttrfig cm]{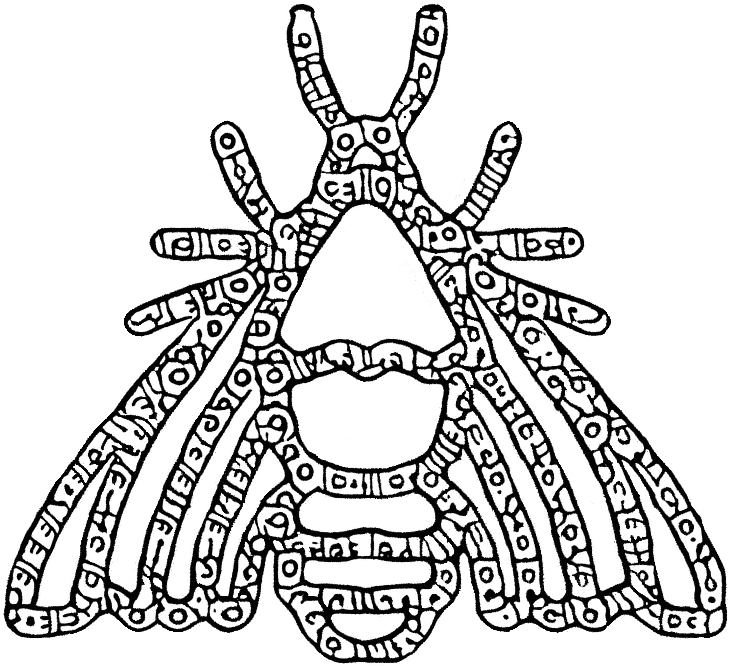}} \\
\multicolumn{2}{c}{(g)} &
\hspace{8pt} &
\multicolumn{2}{c}{(h)}

\end{tabular}

\caption{Qualitative ablation results. Two pairings of sketch and style appear on top, followed by the results of the basic version which employs a reconstruction loss only, in (a) and (b). The results of adding a discriminator follow in (c) and (d), and the combination of reconstruction and shape reconstruction appear in (e) and (f). Our final approach, combining all three losses is shown in (g) and (h), featuring both preservation of content and replication of style. \textit{Skull icon made by \href{https://www.flaticon.com/authors/freepik}{Freepik}, moth sketch by \href{https://www.idangilboa.com/}{Idan Gilboa}.}}
\label{fig:ablation}
\end{figure}

Figure \ref{fig:ablation} features visual examples of the performance of each version. In (a) and (b), we note that the basic version struggles to reconstruct the relevant style elements and generates blurry textures. The second version, shown in (c) and (d), 
suffers from misleading cues given by the discriminator, and resorts to generating empty spaces in place of texture. The third, shown in (e) and (f), like the first, maintains the underlying structure of the sketch, but suffers from style inaccuracies, due to the missing discriminator. Our final solution appears in (g) and (h), and shows an improved ability to synthesize relevant textures, while preserving structure.

\subsection{Comparisons}
\label{sec:comp}

We investigate the potential advantages of arbitrary sketch stylization using our approach vs. two baselines, as well as vs. Image analogies \cite{hertzmann2001image} and two neural style transfer methods: Gatys et al. \shortcite{gatys2016image} and Liao et al. \shortcite{liao2017visual}, both of which employ an optimization process that does not necessitate a large dataset for training.

In these experiments, we make use of typefaces as style exemplars, by rendering 36 glyphs (26 capital letters and 10 digits) and forming a large dataset of patches.
We first train CycleGAN to obtain pairing to plain patches (using plain typefaces), and then proceed to train ST, all as described above.

\begin{figure}[t]
\newcommand{\plfig}{7.5}

\centering
\includegraphics[height=\plfig cm]{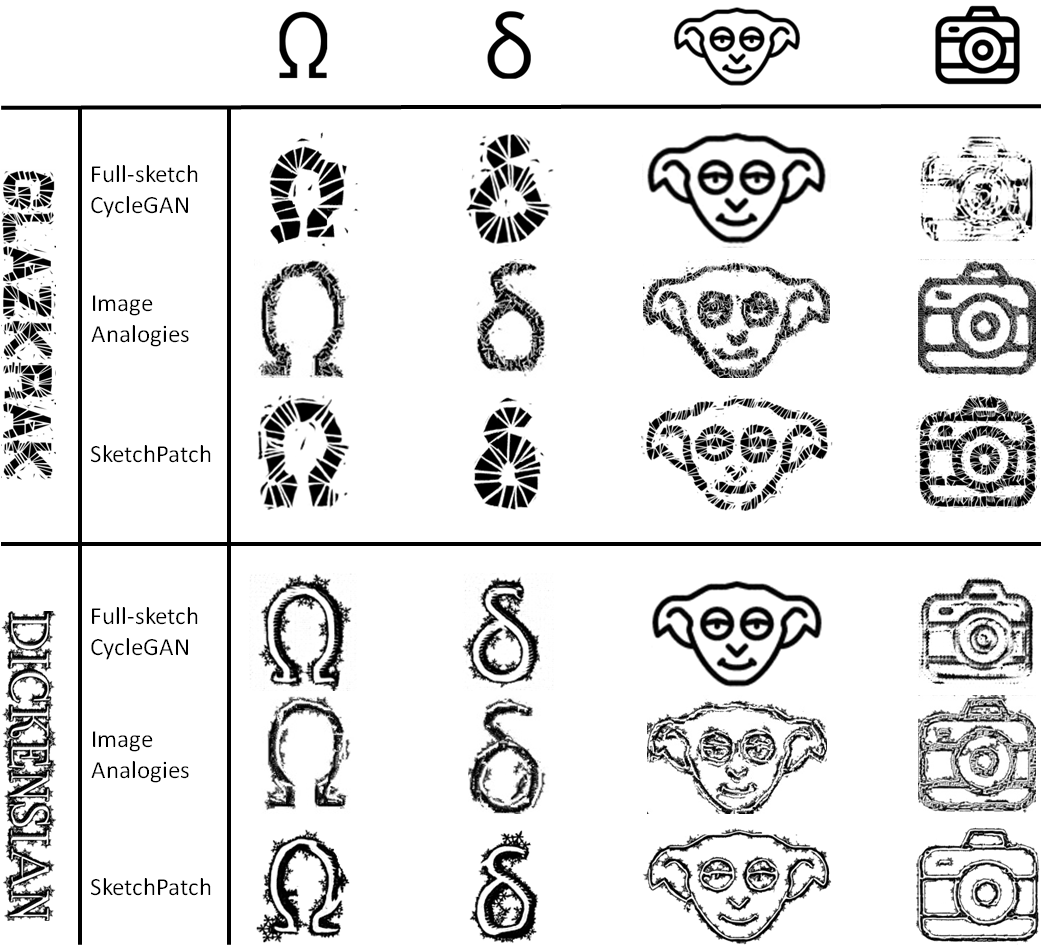}
\caption{Comparison vs. full-sketch CycleGAN baseline (FSCG) and image analogies (IA) on a cross of two styles (far left) with four sketches (top-most). FSCG (rows 1, 4) performs well when given Greek letters as test sketches, since their geometry is similar to its training data. It struggles with more complex geometries, and either produces highly partial and low quality textures (camera), or resorts to taking no action at all (Dobby the house elf).
IA (rows 2, 5) retains the original content, but also struggles with faithfully recreating the patterns. Our method (rows 3, 6) is able to preserve both content and stylistic elements across different styles and sketch geometries.
\textit{Dobby icon by \href{https://www.iconfinder.com/}{Iconfinder}, camera icon by \href{https://www.pngrepo.com/}{PNGRepo}.}}
\label{fig:dovCmp}
\end{figure}

\subsubsection{Baselines}
Our first baseline is essentially our solution for computing pairing when such is unavailable (see Subsection \ref{sec:unpaired}). 
This baseline version is made up of a trained plain-to-styled generator ($P2S$), obtained from training CycleGAN to translate between a pair of styled and plain domains. Given a sketch at test time, we cut it up into patches and translate each of them using the $P2S$ generator. Note that here, the patches are non-overlapping since this approach has no neighborhood considerations. We refer to this baseline as the patch-based CycleGAN baseline (PBCG). 

Our second baseline is also based on CycleGAN, but this time, we train it on full sketches --- glyphs in this instance, rather than on patches. More specifically, the rendered glyphs (36 total) from the styled and plain domains are rotated in increments of one degree to form a dataset of size $\sim$13k. 
We refer to this baseline as the full-sketch CycleGAN baseline (FSCG). 
This version highlights the rigidity involved in learning from a limited set of sketches, and the inevitable inability to generalize to unseen geometries at test time. Figure \ref{fig:dovCmp} provides evidence for this, and shows a set of four sketches translated with this baseline (FSCG), vs. our own method, on two different styles. The sketches contain two Greek letters, and two general sketches with more complex geometries than the letters. As can be seen, FSCG performs well on the test letters, since their geometry, while not identical, is very similar to the geometry of the sketches it was trained on. However, when asked to translate more complex sketches, it struggles, and either produces highly partial and low quality texture (fourth column), or defaults to the original sketch without altering it (third column). This baseline is not included in our main comparison experiment, due to its inability to produce relevant comparable content for complex sketch geometries.

This comparison also includes results obtained by activating Image analogies on the test sketches. This method receives a trio of images, such that the first two demonstrate the expected transformation, and the third provides the "canvas" upon which to apply the learned transformation. We prepared two pairs of images, one for each presented style. Each pair contains a plain image and a styled image to demonstrate the transformation, and is composed of a grid of aligned glyphs. Since we do not possess the pairing between the plain and styled glyphs, we used FSCG to generate a corresponding plain glyph per styled one. We note that this method preserves the content of the test sketch as well as some of the stylistic elements, but some are not transferred successfully onto the new geometry.

\begin{figure}
\newcommand{\ttrfig}{4}
\newcommand{\ttbfig}{5}
\setlength\tabcolsep{1pt}

\centering
\begin{tabular}[t]{c c}

\includegraphics[height=\ttrfig cm]{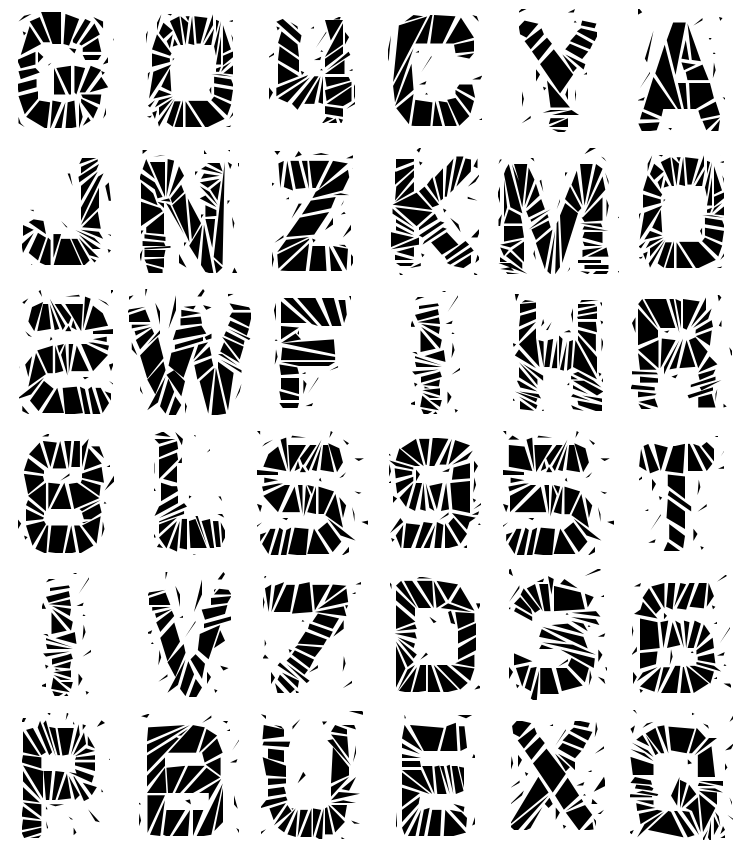} \hspace{8pt} &
\includegraphics[height=\ttrfig cm]{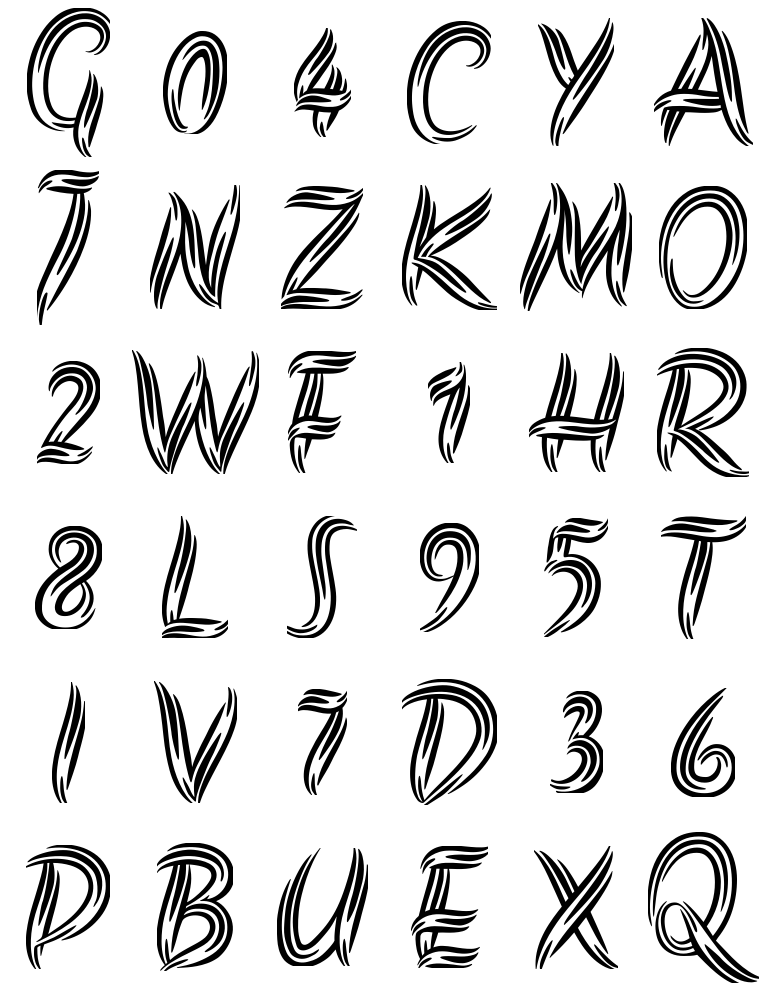} \hspace{8pt} \\
(a) \hspace{8pt} & (b) \\

\end{tabular}

\caption{Style grid examples for neural style transfer. In order to compare to neural style transfer methods, the 36 glyphs of a given style were arranged in a 6x6 grid to form a style exemplar. The font \textit{Glazkrak} is in (a), and \textit{Akronim} in (b).}
\label{fig:st_style_imgs}
\end{figure}

\begin{figure*}[h]
\newcommand{\figOrig}{1.5}
\newcommand{\figStyle}{1}

\setlength\tabcolsep{1pt}
\centering

\newcolumntype{Y}{>{\centering\arraybackslash}X}

\begin{tabularx}{16cm}{c *{10}{Y}}
\includegraphics[height=\figOrig cm]{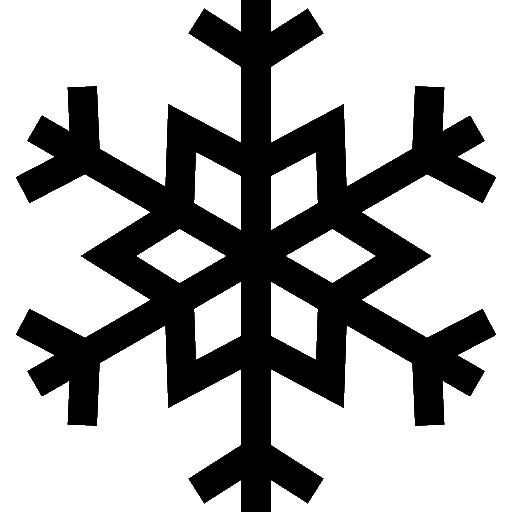}&
\includegraphics[height=\figOrig cm]{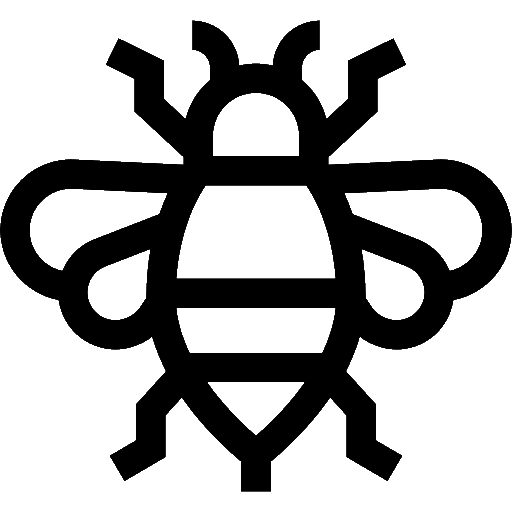} &
\includegraphics[height=\figOrig cm]{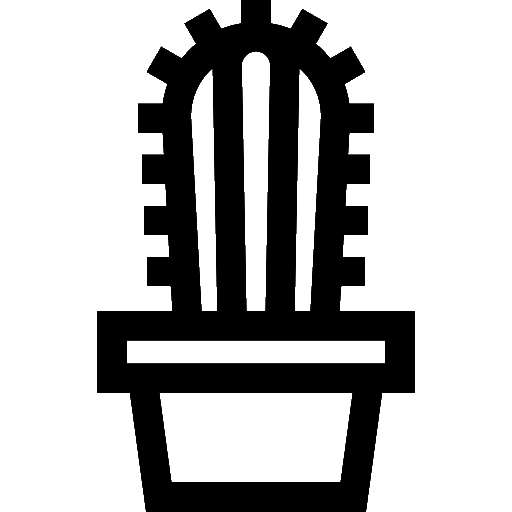} &
\includegraphics[height=\figOrig cm]{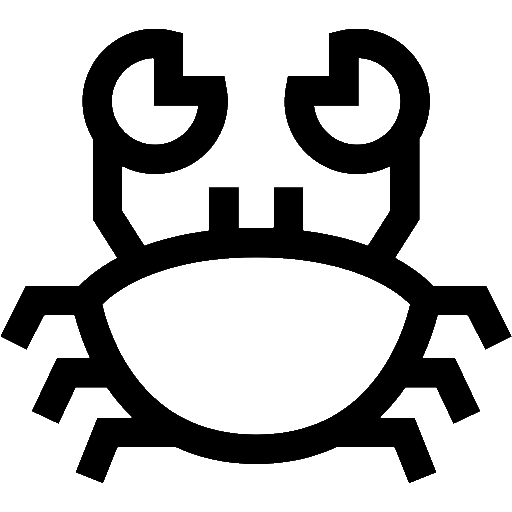}&
\includegraphics[height=\figOrig cm]{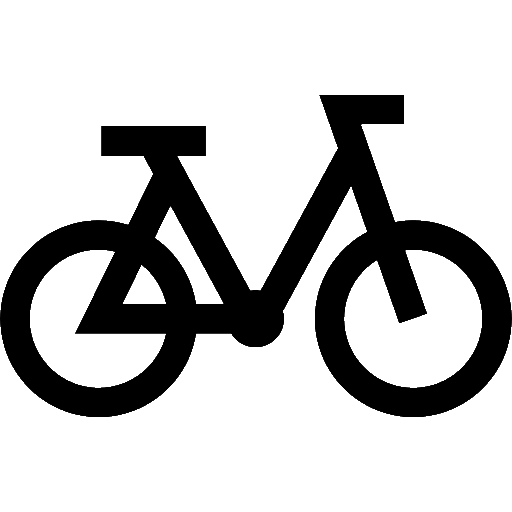}&
\includegraphics[height=\figOrig cm]{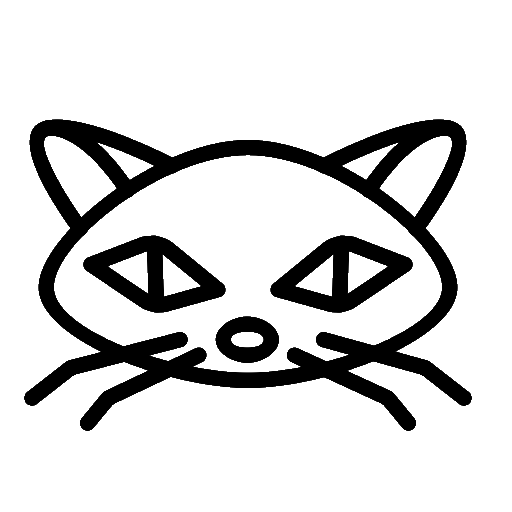}&
\includegraphics[height=\figOrig cm]{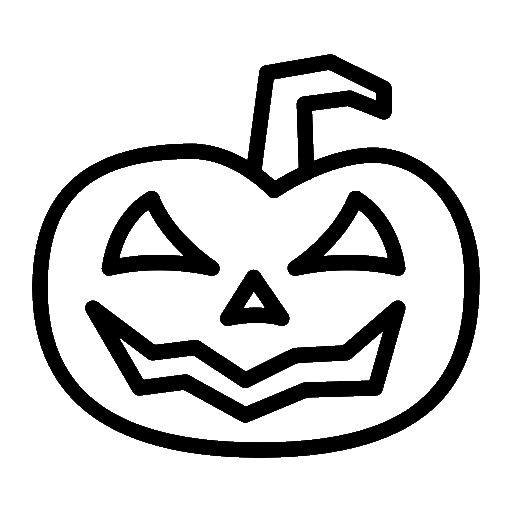}&
\includegraphics[height=\figOrig cm]{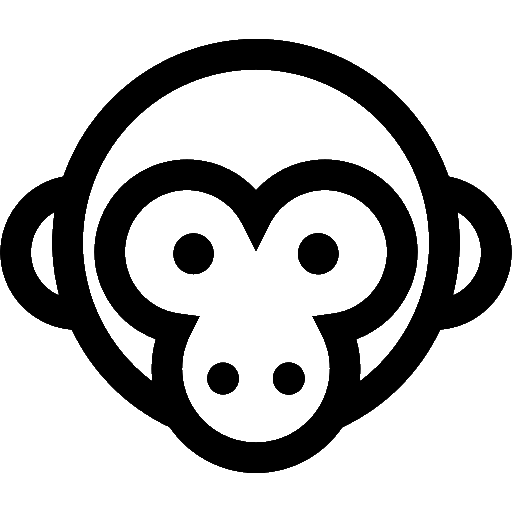} &
\includegraphics[height=\figOrig cm]{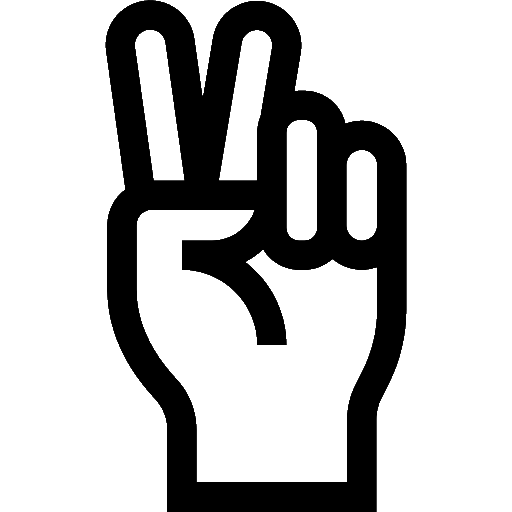} &
\includegraphics[height=\figOrig cm]{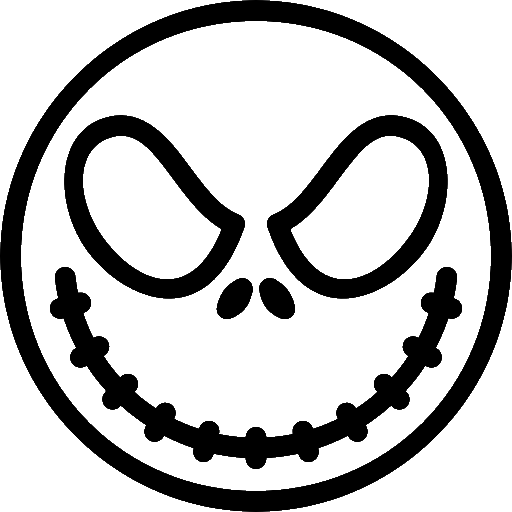} \\
\multicolumn{10}{c}{(a)}\\
\end{tabularx}

\vspace{0.5cm}

\begin{tabularx}{16cm}{c *{5}{Y}}

\includegraphics[height=\figStyle cm]{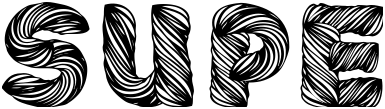} &
\includegraphics[height=\figStyle cm]{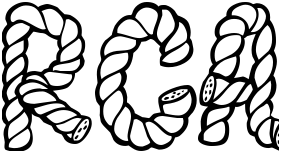} &

\includegraphics[height=\figStyle cm]{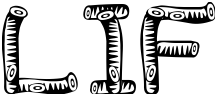} &
\includegraphics[height=\figStyle cm]{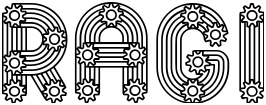} &
\includegraphics[height=\figStyle cm]{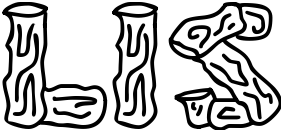}\\

\includegraphics[height=\figStyle cm]{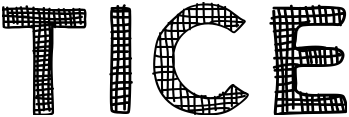}&
\includegraphics[height=\figStyle cm]{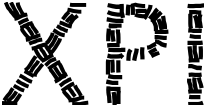} &
\includegraphics[height=\figStyle cm]{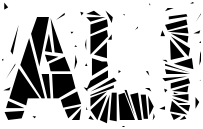} &
\includegraphics[height=\figStyle cm]{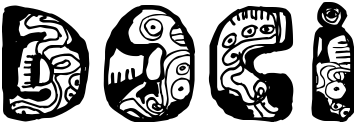} &
\includegraphics[height=\figStyle cm]{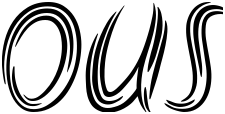}\\
\multicolumn{5}{c}{}\\
\multicolumn{5}{c}{(b)}\\

\end{tabularx}

\caption{Test set for comparisons. We designated 10 sketches (a), and 10 styles (b) for our comparisons.
\textit{Snowflake, bee, cactus, crab, bicycle, monkey and hand icons made by \href{https://www.flaticon.com/authors/freepik}{Freepik}; cat, pumpkin and Jack Skellington icons by \href{https://www.iconfinder.com/}{Iconfinder}.}
}
\label{fig:testTimeInput}
\end{figure*}

\subsubsection{Style transfer}
In order to compare to Gatys et al. \shortcite{gatys2016image} and Liao et al. \shortcite{liao2017visual}, we prepare a style image for each of our test set styles. Here, too, style exemplars are taken from decorative typefaces. We place all the glyphs of a selected typeface within a 6x6 grid 
(see Figure \ref{fig:st_style_imgs} for an example). 
Each of these two methods is then run with input pairs of images, where one is taken from our test set sketches, and the other is our prepared grid style image.

\subsubsection{Comparison discussion}
\begin{figure*}[h]

\newcommand{\figOrig}{3}
\newcommand{\figCatDia}{2.8}
\newcommand{\figCycling}{2.9}

\newcommand{\ablfigsp}{0.2} 
\setlength\tabcolsep{1pt}
\centering

\begin{tabular}[t]{ c c c c}

\includegraphics[height=\figOrig cm]{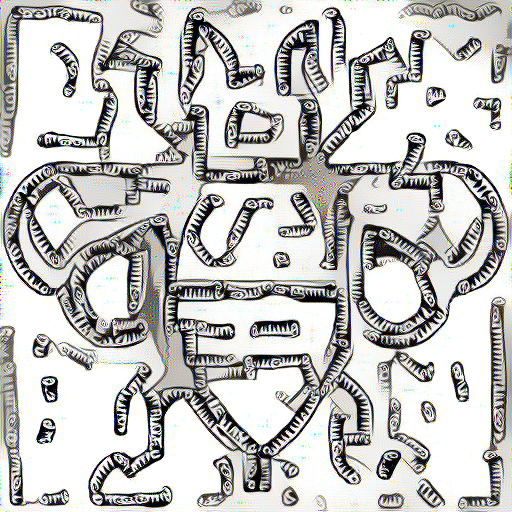}&
\includegraphics[height=\figOrig cm]{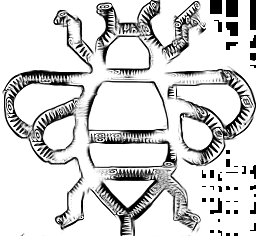} &
\includegraphics[height=\figOrig cm]{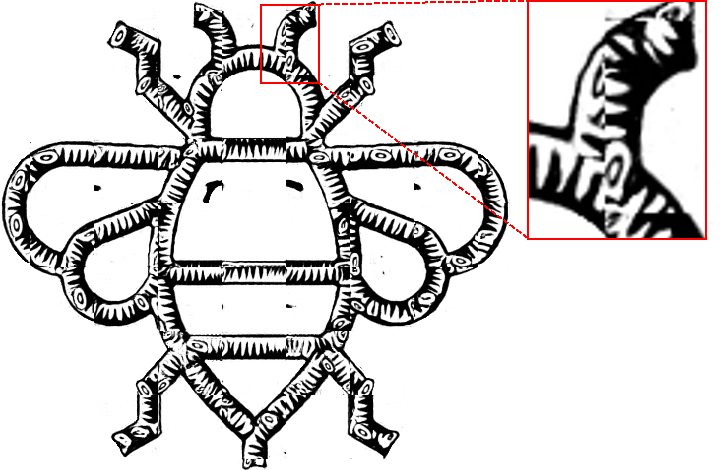}&
\includegraphics[height=\figOrig cm]{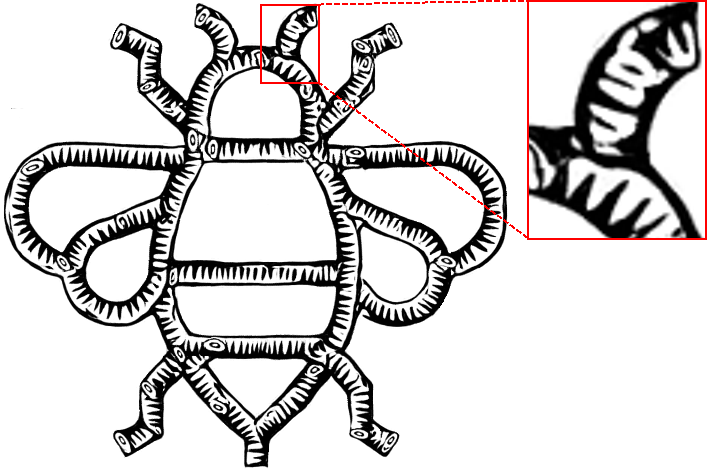} \\

\includegraphics[height=\figOrig cm]{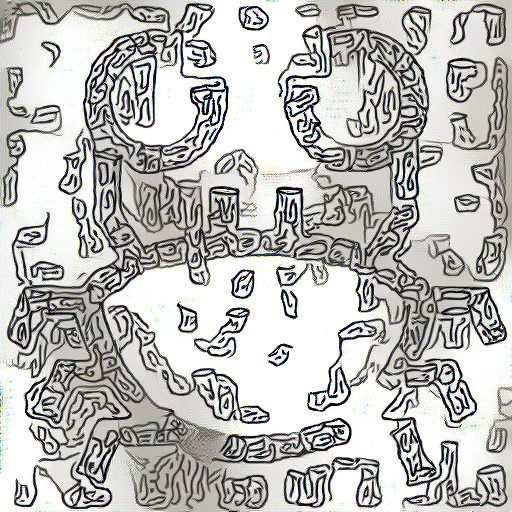}&
\includegraphics[height=\figOrig cm]{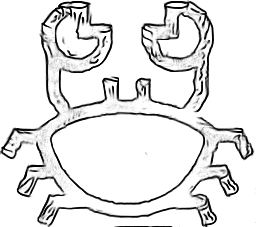} &
\includegraphics[height=\figOrig cm]{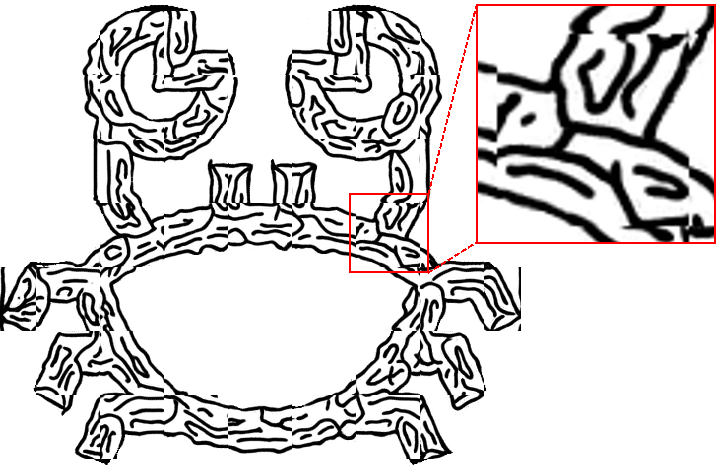}&
\includegraphics[height=\figOrig cm]{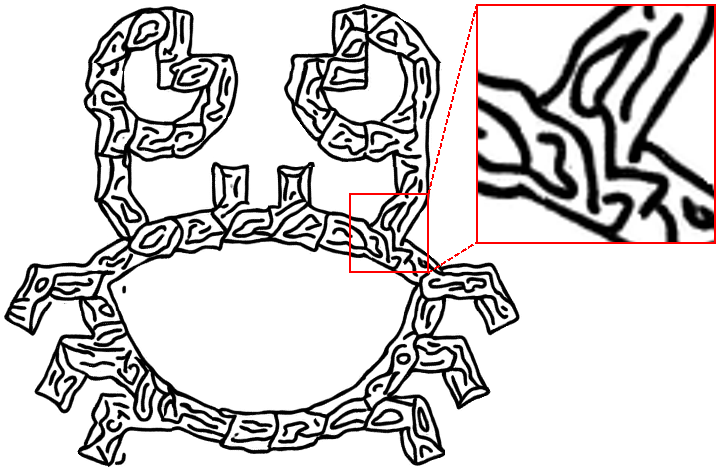} \\

\includegraphics[height=\figOrig cm]{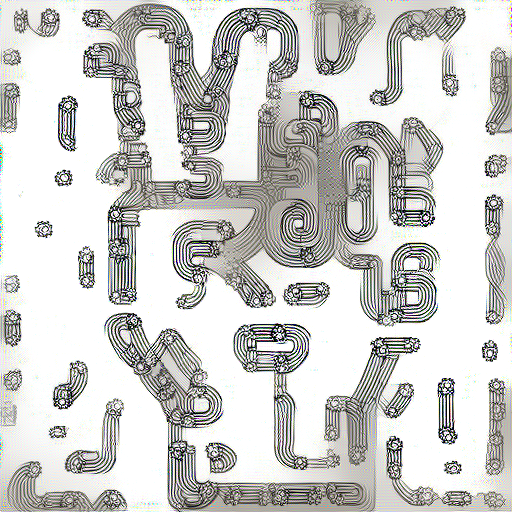}&
\includegraphics[height=\figOrig cm]{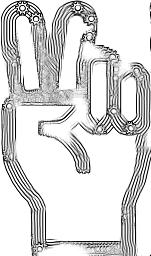} &
\includegraphics[height=\figOrig cm]{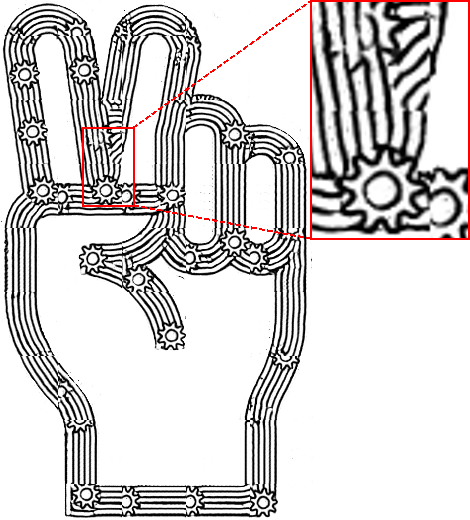}&
\includegraphics[height=\figOrig cm]{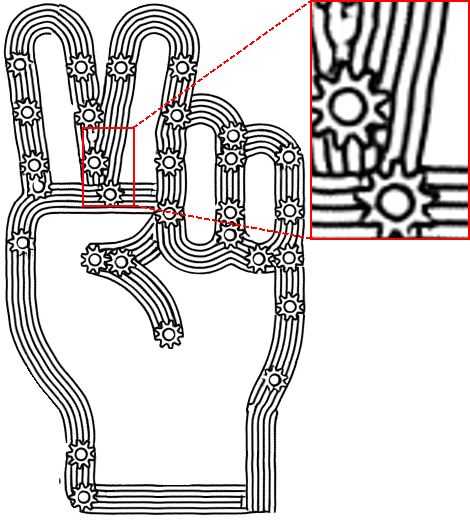} \\

\includegraphics[height=\figOrig cm]{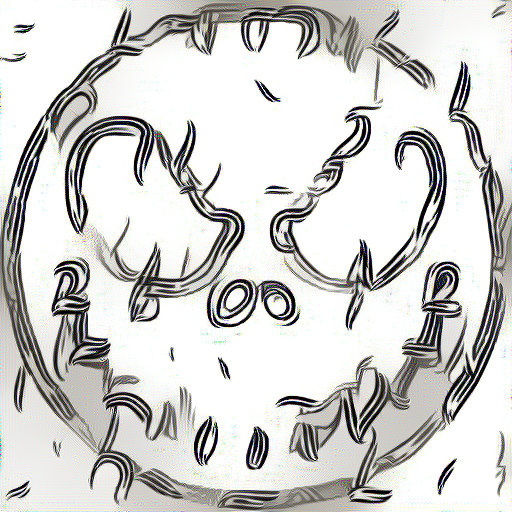}&
\includegraphics[height=\figOrig cm]{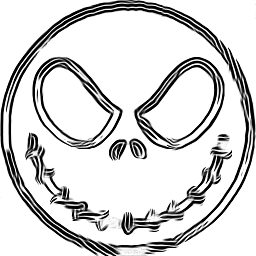} &
\includegraphics[height=\figOrig cm]{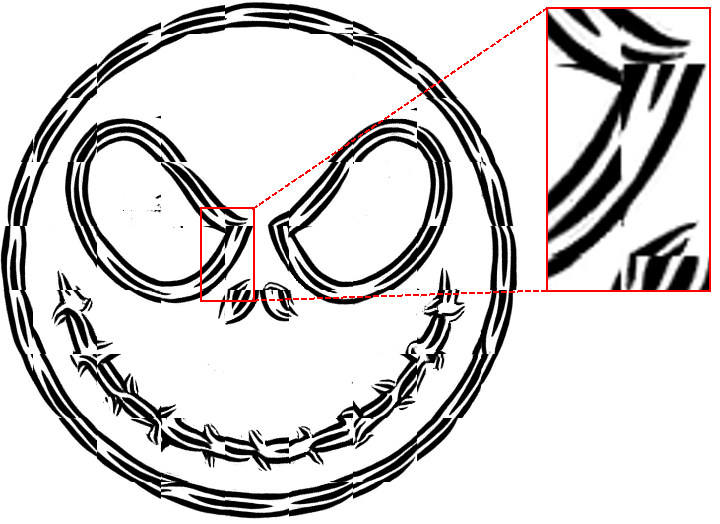}&
\includegraphics[height=\figOrig cm]{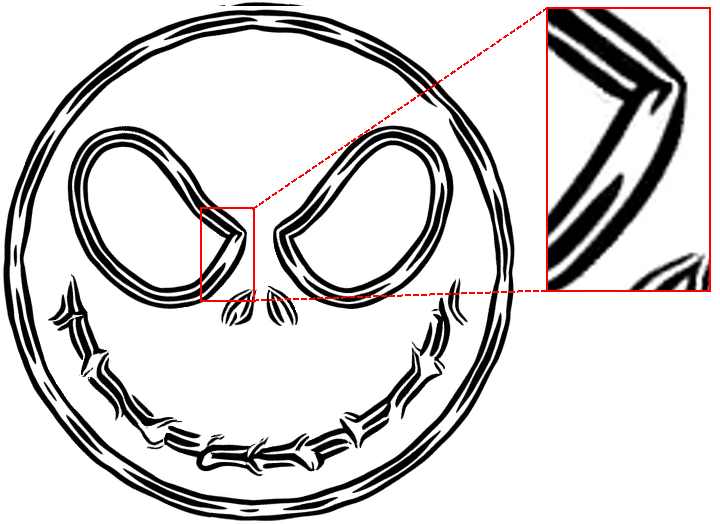} \\

\includegraphics[height=\figOrig cm]{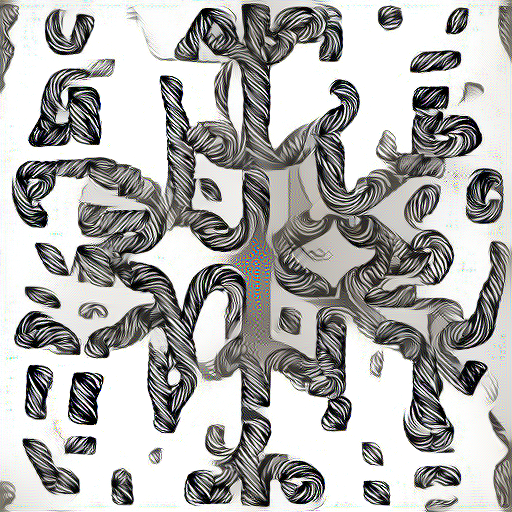}&
\includegraphics[height=\figOrig cm]{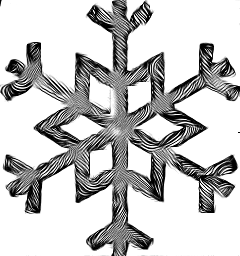} &
\includegraphics[height=\figOrig cm]{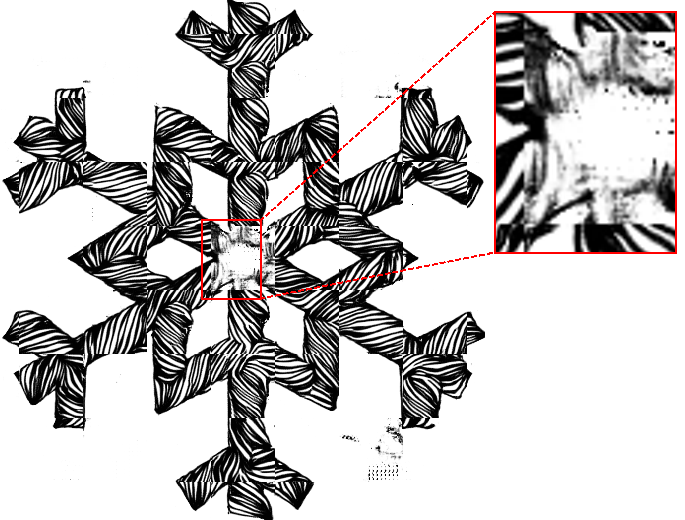}&
\includegraphics[height=\figOrig cm]{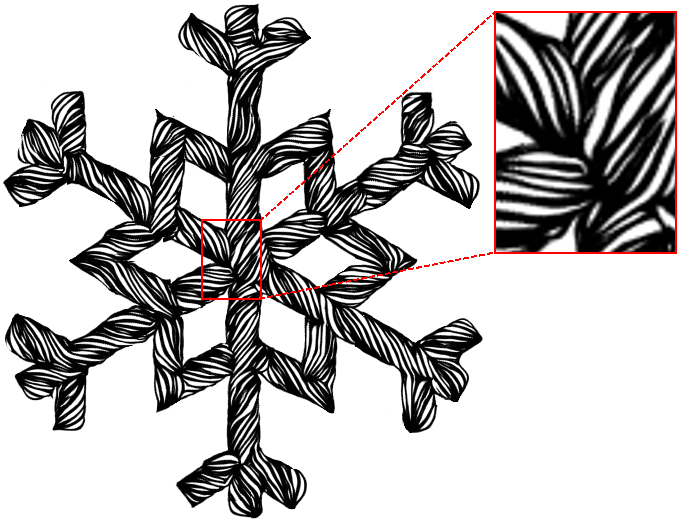} \\

\includegraphics[height=\figOrig cm]{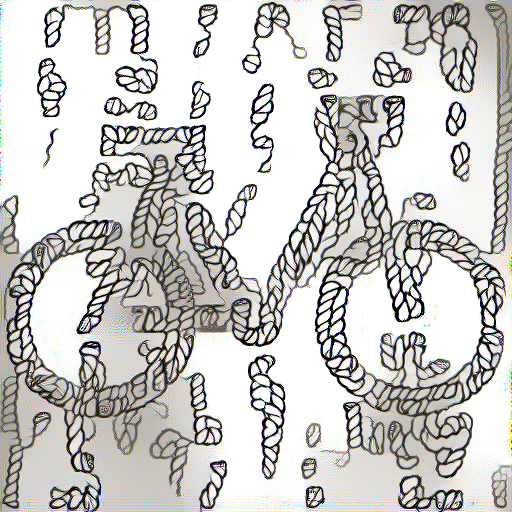}&
\includegraphics[height=\figCycling cm]{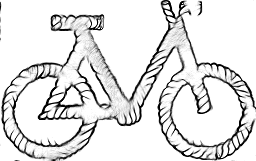} &
\includegraphics[height=\figCycling cm]{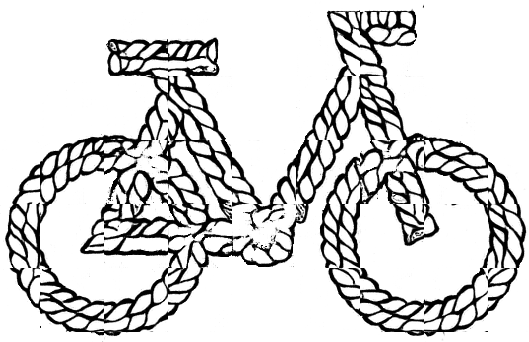}&
\includegraphics[height=\figCycling cm]{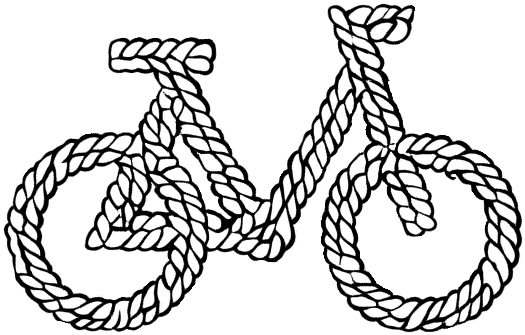} \\

\vspace{1pt}\\

Gatys et al.[2016] &
Liao et al.[2017] &
Patch-based CycleGAN &
Ours \\

\end{tabular}

\caption{Comparison results on the test set shown in Figure \ref{fig:testTimeInput}. The results of Gatys et al. \shortcite{gatys2016image} appear in the first column, followed by Liao et al. \shortcite{liao2017visual} in the second. Next, our patch-based CycleGAN baseline (PBCG) appears in the third column, and our approach in the fourth.}
\label{fig:comparison}
\end{figure*}

Figure \ref{fig:comparison} presents a set of five sketches featuring five different styles (all from our designated test set in Figure \ref{fig:testTimeInput}), by each of the four compared methods. 
In the first column, we note that Gatys et al. \shortcite{gatys2016image} mostly succeed in capturing the style elements in the exemplar given to it as the style image (see Figure \ref{fig:st_style_imgs}), but is unable to fully separate the content from the style, resulting in scattered artifacts throughout the image.
In the second column, Liao et al. \shortcite{liao2017visual} preserve the content well by matching patches while optimizing and operating on multiple scales of deep feature maps, but the style elements are not consistently reconstructed, particularly, it seems, in regions that are more geometrically complex, such that their liking could not be found in the style exemplar. Note that this comparison is somewhat inadequate, since this method is not geared toward such a setting, but rather aims to synthesize analogies of two images that share semantic similarities. Our patch-based CycleGAN baseline appears in the third column, displaying content preservation alongside faithful style synthesis, but upon closer scrutinization, one may discover pattern breakage and discontinuities along patch borders, as can be seen in the highlighted insets within the figure. Finally, our approach is shown in the fourth column, combining the benefits of high quality patch-based translation, with seamless transitions between patches. Please refer to our supplementary material for more comparisons.

\subsection{Results}
\label{sec:results}

\begin{figure}
\newcommand{\ttrfig}{3.7}
\newcommand{\ttsfig}{1.2}
\setlength\tabcolsep{1pt}

\centering
\begin{tabular}[t]{c c c c c}

\includegraphics[height=\ttsfig cm]{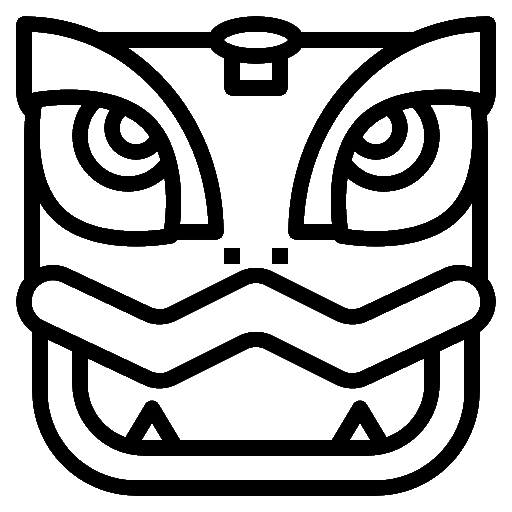} &
\includegraphics[height=\ttsfig cm]{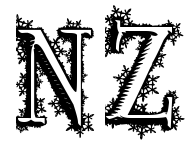} &
\hspace{8pt} &
\includegraphics[height=\ttsfig cm]{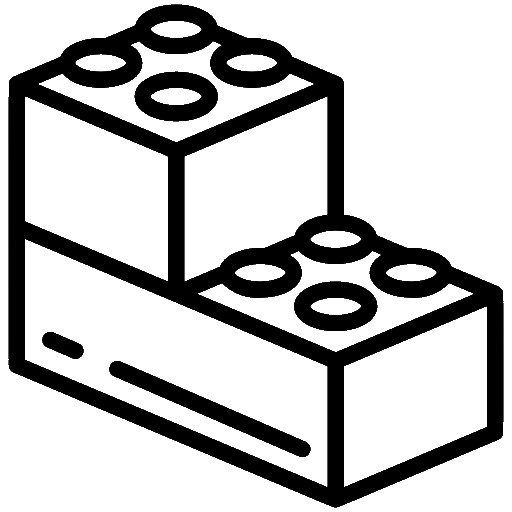} &
\includegraphics[height=\ttsfig cm]{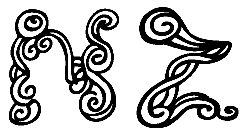} \\
 
\hline
\\
\multicolumn{2}{c}
{\includegraphics[height=\ttrfig cm]{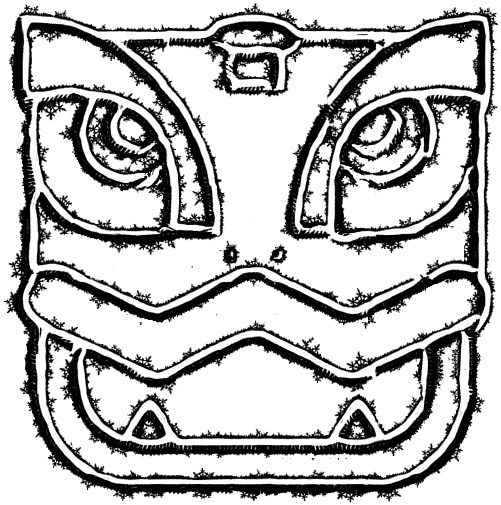}} &
\hspace{8pt}  &
\multicolumn{2}{c}
{\includegraphics[height=\ttrfig cm]{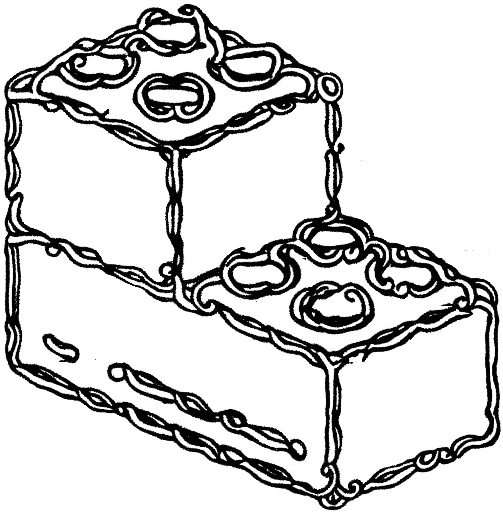}} \\
\multicolumn{2}{c}{(a)} &
\hspace{8pt} &
\multicolumn{2}{c}{(b)} \\
\multicolumn{2}{c}
{\includegraphics[height=\ttrfig cm]{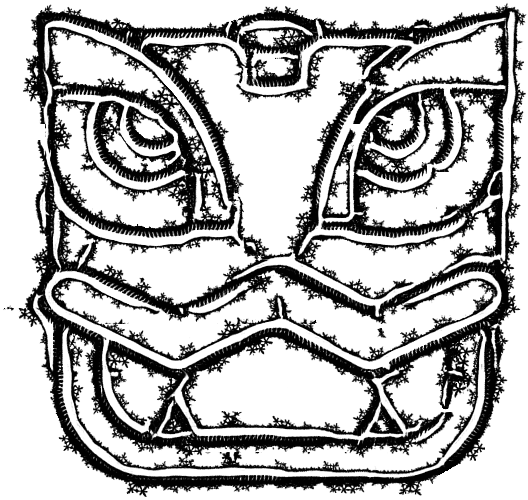}} &
\hspace{8pt} &
\multicolumn{2}{c}
{\includegraphics[height=\ttrfig cm]{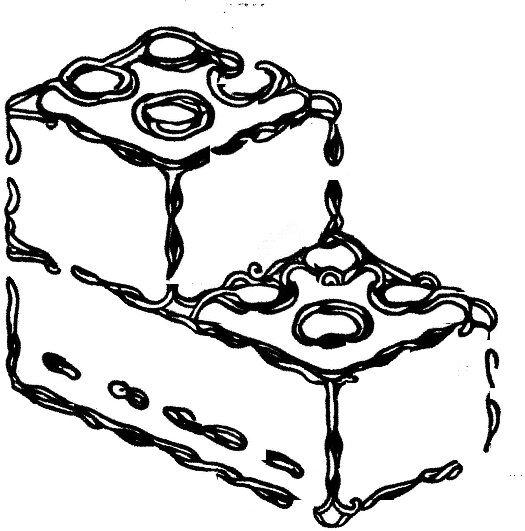}} \\
\multicolumn{2}{c}{(c)} &
\hspace{8pt} &
\multicolumn{2}{c}{(d)} \\

\end{tabular}
\caption{Plain domain impact. The two featured styles were each trained twice -- once with a plain font of a similar weight, and once with a dataset composed of patches of plain sketches, of an appropriate stroke weight as well. Results obtained by training against a plain font generally preserve the shape better (a,b) than those obtained using a general pool of primitives (c,d). 
\textit{Icons made by \href{https://www.flaticon.com/authors/freepik}{Freepik}.}
}

\label{fig:plain_domain}
\end{figure}

Apart from the results shown in the previous section, we include additional experiments to further evaluate the performance of our solution. A gallery of results can be found in Figures \ref{fig:charachter_parade} and \ref{fig:naive_parade}. A large body of extra results appears in the supplementary material.

\subsubsection{Plain domain impact}
For unpaired style exemplars,
we explore the impact of the nature of the plain domain that we choose to match to it.
This experiment is conducted using typefaces, due to the
inherent geometric compatibility among different individuals in this realm (all feature highly similar sets of geometric shapes --- glyphs). This compatibility sets the stage for an interesting test, where we first enlist a geometrically similar plain typeface to provide plain domain patches for a given styled domain, with which to procure the necessary pairing. 
Despite the absence of an explicit pairing between the primitives in the styled domain, with those in the plain domain, there exists a strong geometric connection linking the two, stemming simply from their shared typographical origin. 
Second, we deliberately ignore this inherent compatibility, and select a pool of miscellaneous plain sketches to provide plain patches for pairing, and examine the impact that such a pairing can have upon the final result. For this experiment, we collect two datasets containing patches extracted from plain sketches (similar to the ones in our test set), such that one features thin strokes, and the other thick ones, and the stroke weight of the styled typeface determines which of the two is used as the plain domain.

In both experiments, patch-based CycleGAN is used to generate an aligned pairing for each styled patch as explained in Subsection \ref{sec:unpaired}. After generating the pairing, we train ST, and subsequently apply the trained model to obtain test results.

Figure \ref{fig:plain_domain} presents these results, where the images in (a) and (b) were obtained using a plain typeface for pairing, and those in (c) and (d) were obtained using the sketch-based datasets for pairing. It is unsurprising to note that the underlying geometry of the shapes is better preserved when training against primitives featuring similar geometries, since CycleGAN is more likely to learn an adequate translation. Observe, for instance, the dragon in (c), and the strokes that were added to the tip of its bottom teeth. These strokes are reminiscent of a serif, and seeing as in this instance, the plain domain lacks this particular type of geometric feature, the network learns to add it to certain strokes.

\subsubsection{Artist and novice user designed styles}
\begin{figure}
\newcommand{\ttrfig}{4.8}
\newcommand{\ttsfig}{1.2}
\setlength\tabcolsep{1pt}

\centering
\begin{tabular}[t]{c c}

\adjincludegraphics[height=\ttrfig cm,trim={0 {.15\height} 0 0},clip]{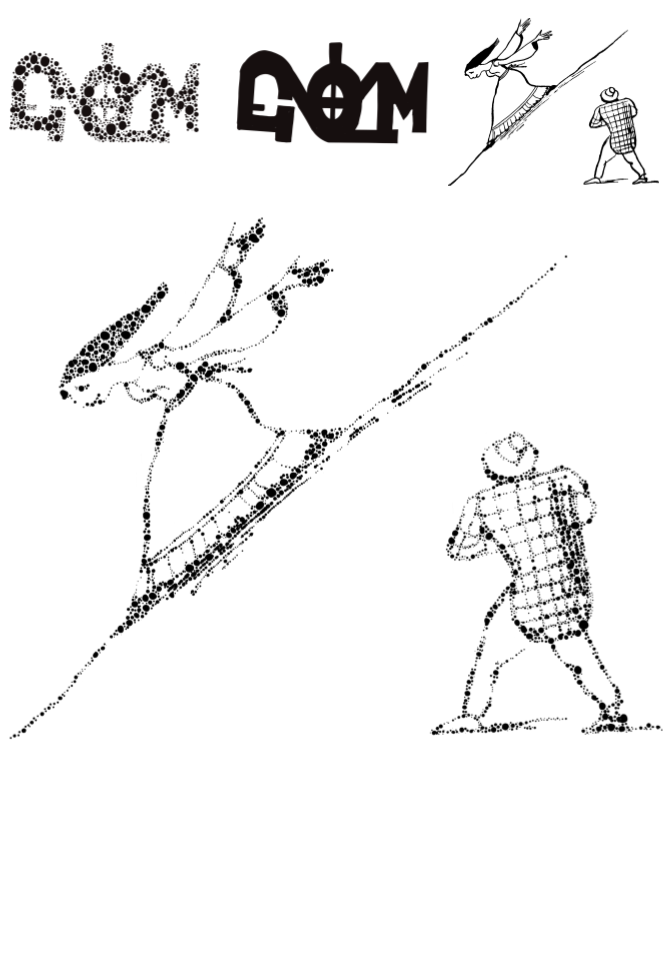} &
\adjincludegraphics[height=\ttrfig cm,trim={0 {.1\height} 0 0},clip]{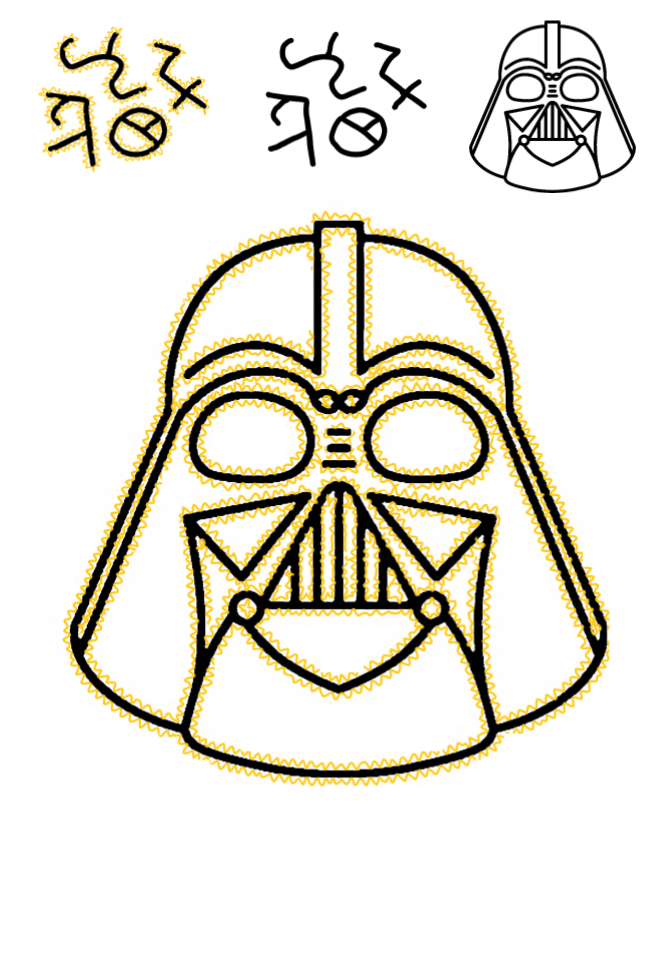} \\
\adjincludegraphics[height=\ttrfig cm,trim={0 {.25\height} 0 0},clip]{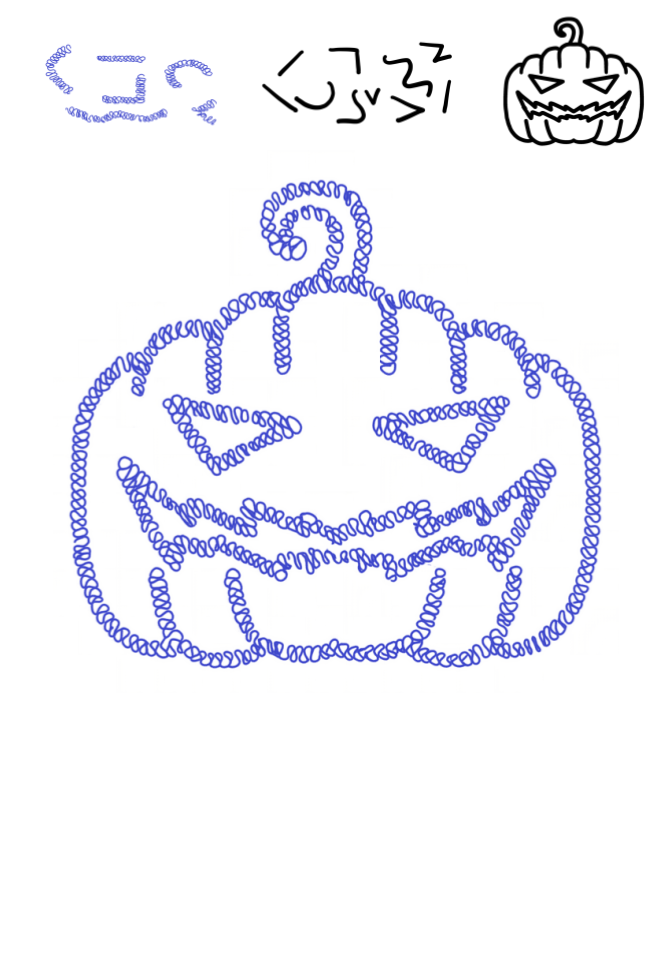} &
\adjincludegraphics[height=\ttrfig cm,trim={0 {.15\height} 0 0},clip]{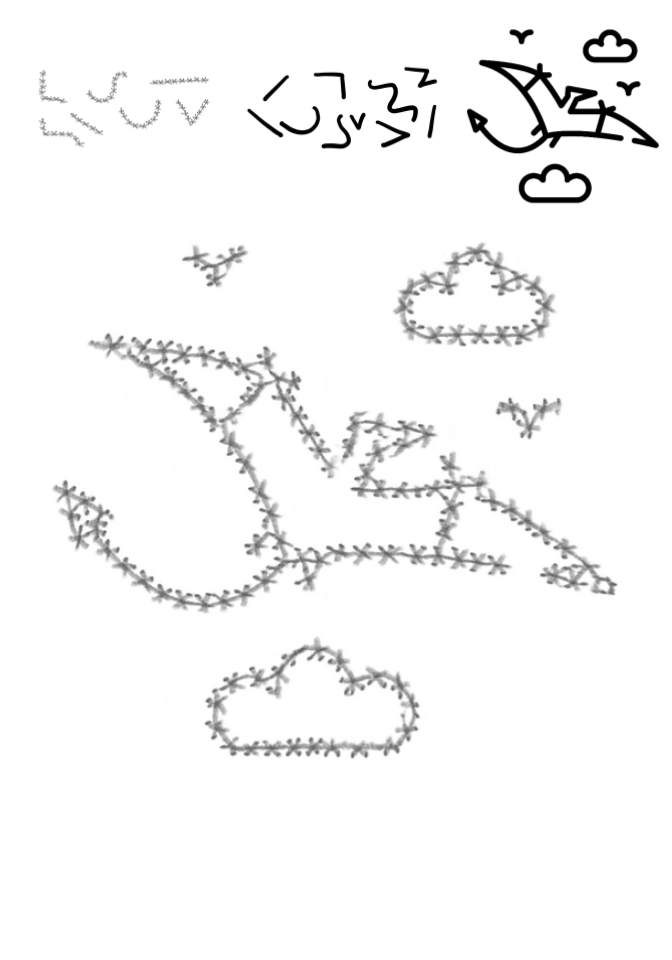} \\

\end{tabular}

\caption{Artist and novice user designed styles. 
Each of the four examples features a style exemplar, its plain counterpart and the test sketch, up top, and the resulting styled sketch as the center piece. An artist designed style exemplar that is pre-paired with an aligned plain exemplar appears at the top-left.
Another paired exemplar, this time by a novice user, appears at the top-right. The bottom row contains two exemplars by a novice, both of which were matched with the same plain stroke library for pair alignment.
\textit{
Vintage sketch from the Book of Limericks 1888,
Darth Vader icon by \href{https://www.pngrepo.com/}{PNGRepo},
pumpkin sketch by \href{https://www.idangilboa.com/}{Idan Gilboa},
dinosaur icon by \href{https://www.flaticon.com/authors/freepik}{Freepik}.}
}
\label{fig:artist}
\end{figure}
In this experiment, we obtain a number of style exemplars designed by an artist as well as a novice user. 
Both used a pen tablet and were asked to produce a styled sketch as well as a plain one. In some, the plain one was produced as the backbone for the styled counterpart, such that they are perfectly aligned. In others, it was created as a post-process, where the user was asked to simply draw plain strokes that are geometrically similar, but not identical, to those they produced for their styled exemplar. The first variety yields paired patches, and was therefore trained using the paired pipeline. The other is unpaired, and was trained using the unpaired pipeline consisting of training CycleGAN to compute the pairing, and then proceeded to the paired pipeline of ST.

Figure \ref{fig:artist} showcases four different style exemplars and the result of applying the ST model trained on each one. Note that despite the limited variety of strokes contained in these exemplars, ST is able to generalize well to arbitrary geometries at inference time. Moreover, the bottom two exemplars contain no intersections at all, yet the resulting styled sketches remain true-to-style, even across diverging structures. However, these are not highly intricate patterns, which is a helping factor in this case.
More results obtained using these exemplars can be found in our supplementary material.

\subsubsection{Natural media stylization}
\begin{figure}
\newcommand{\ttrfig}{3.7}
\newcommand{\ttsfig}{1.2}
\setlength\tabcolsep{1pt}

\centering
\begin{tabular}[t]{c c c c c}

\includegraphics[height=\ttsfig cm]{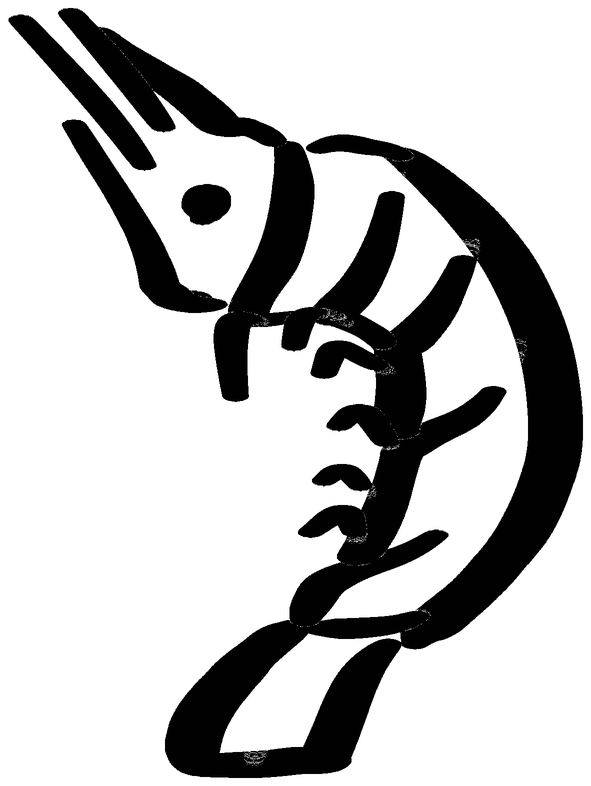} &
\includegraphics[height=\ttsfig cm]{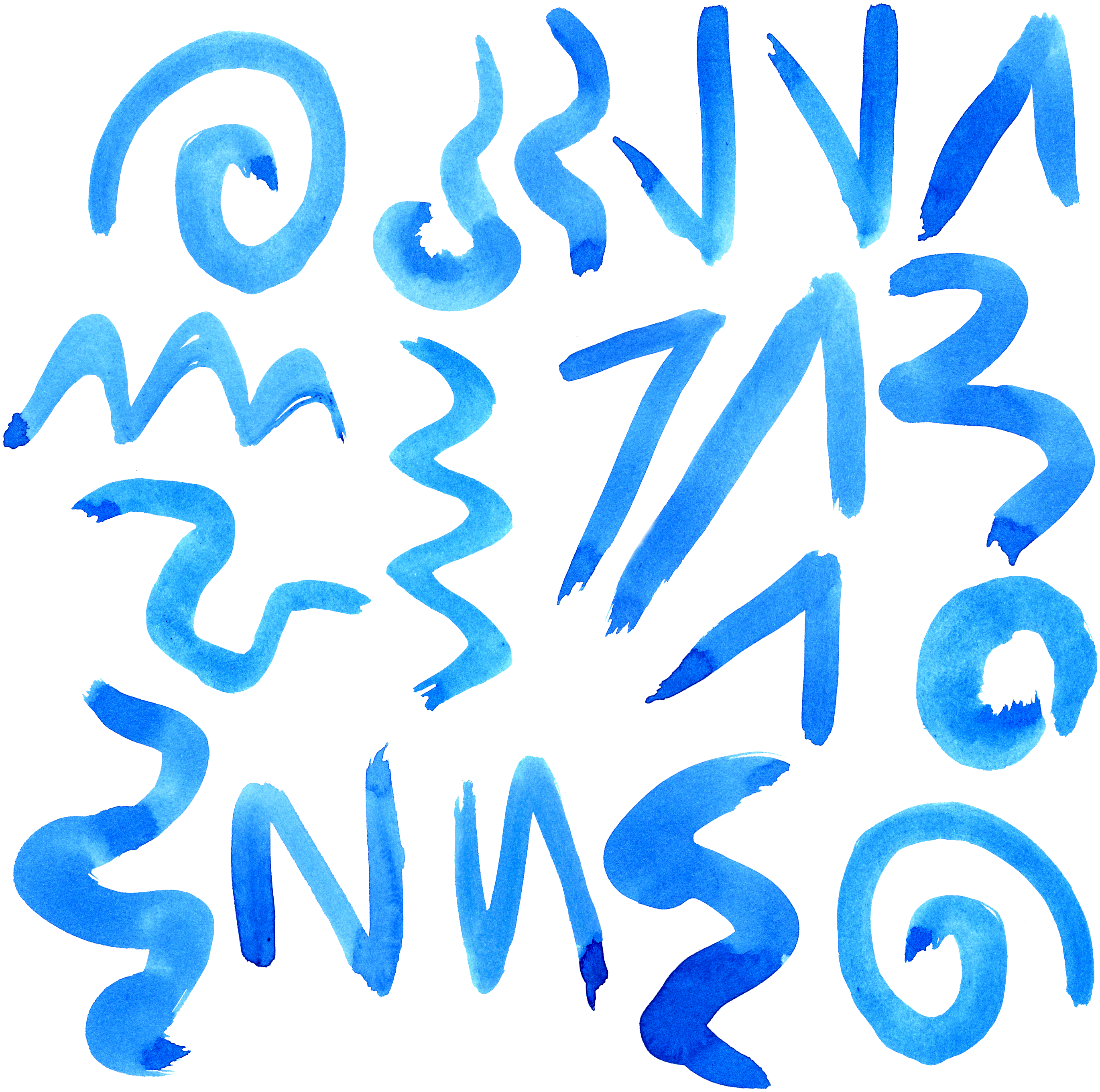} &
\hspace{8pt} &
\includegraphics[height=\ttsfig cm]{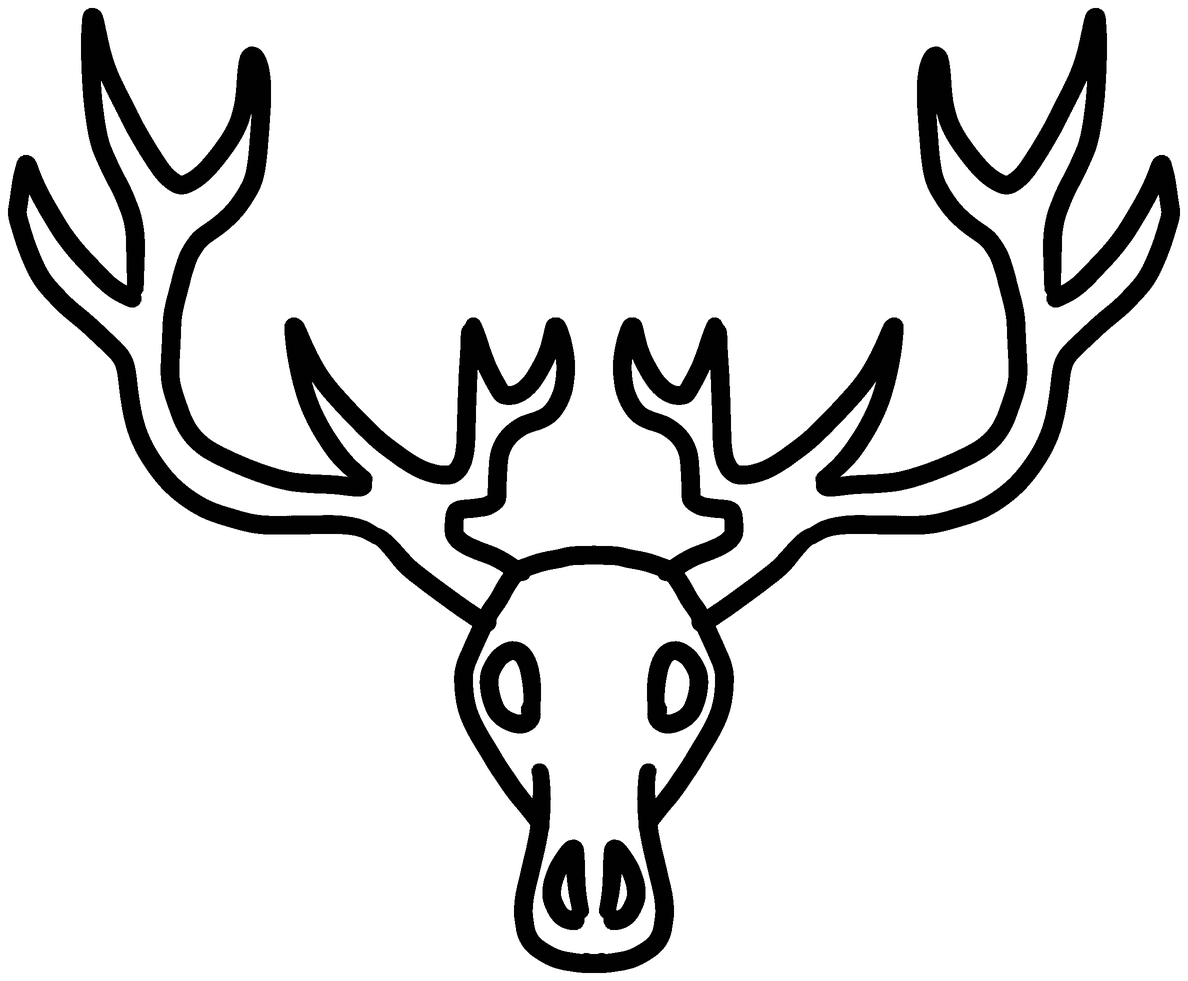} &
\includegraphics[height=\ttsfig cm]{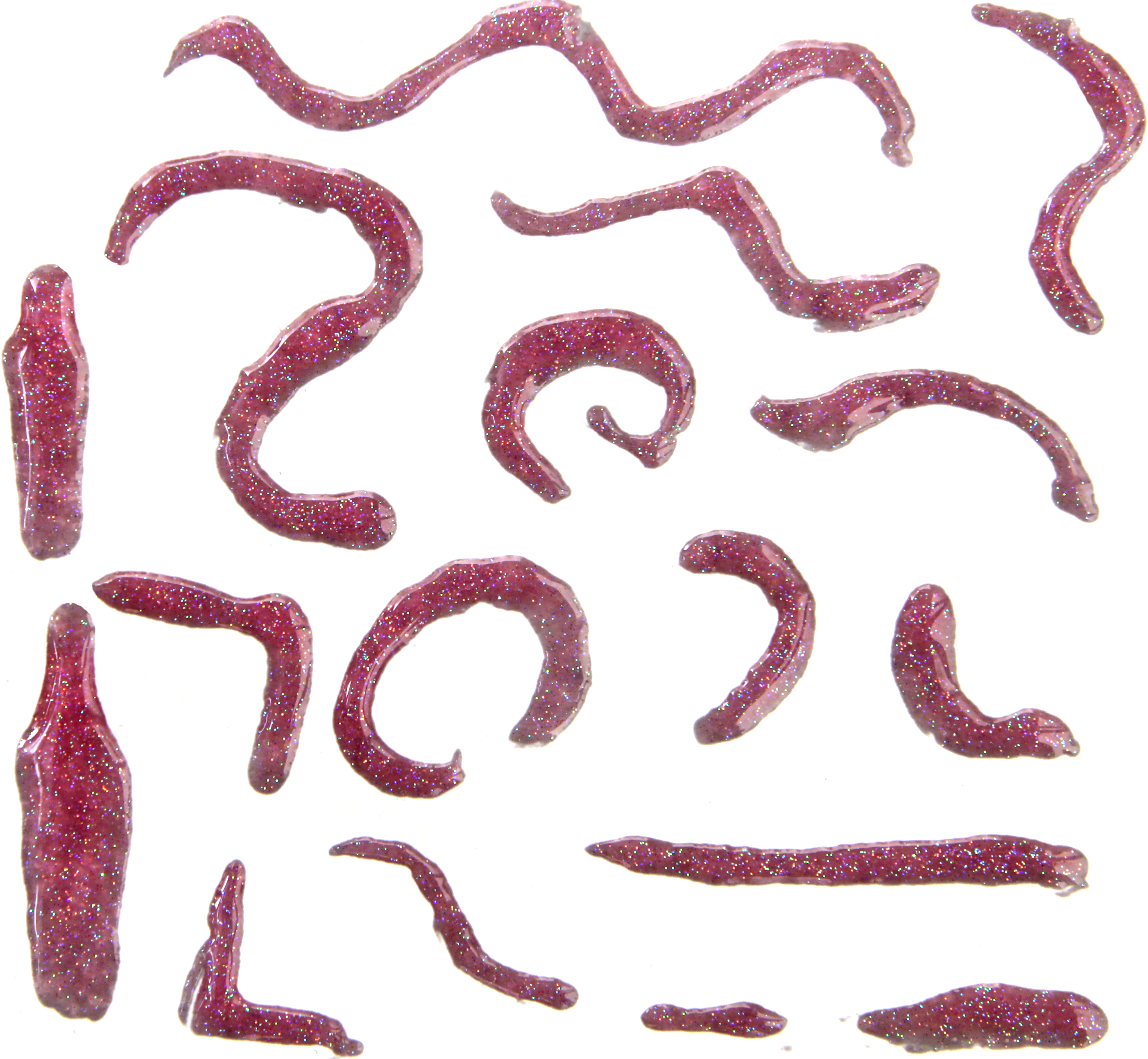}  \\
\hline
\\
\multicolumn{2}{c}
{\includegraphics[height=\ttrfig cm]{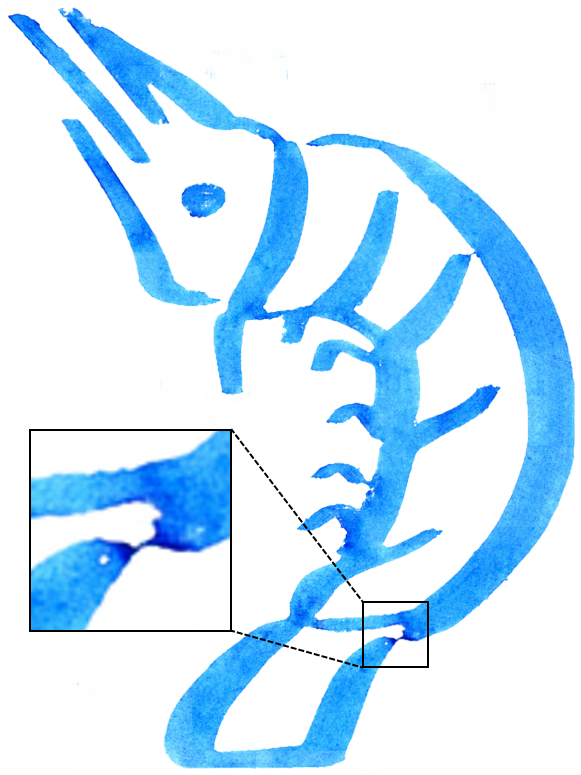}} &
\hspace{8pt}  &
\multicolumn{2}{c}
{\includegraphics[height=\ttrfig cm]{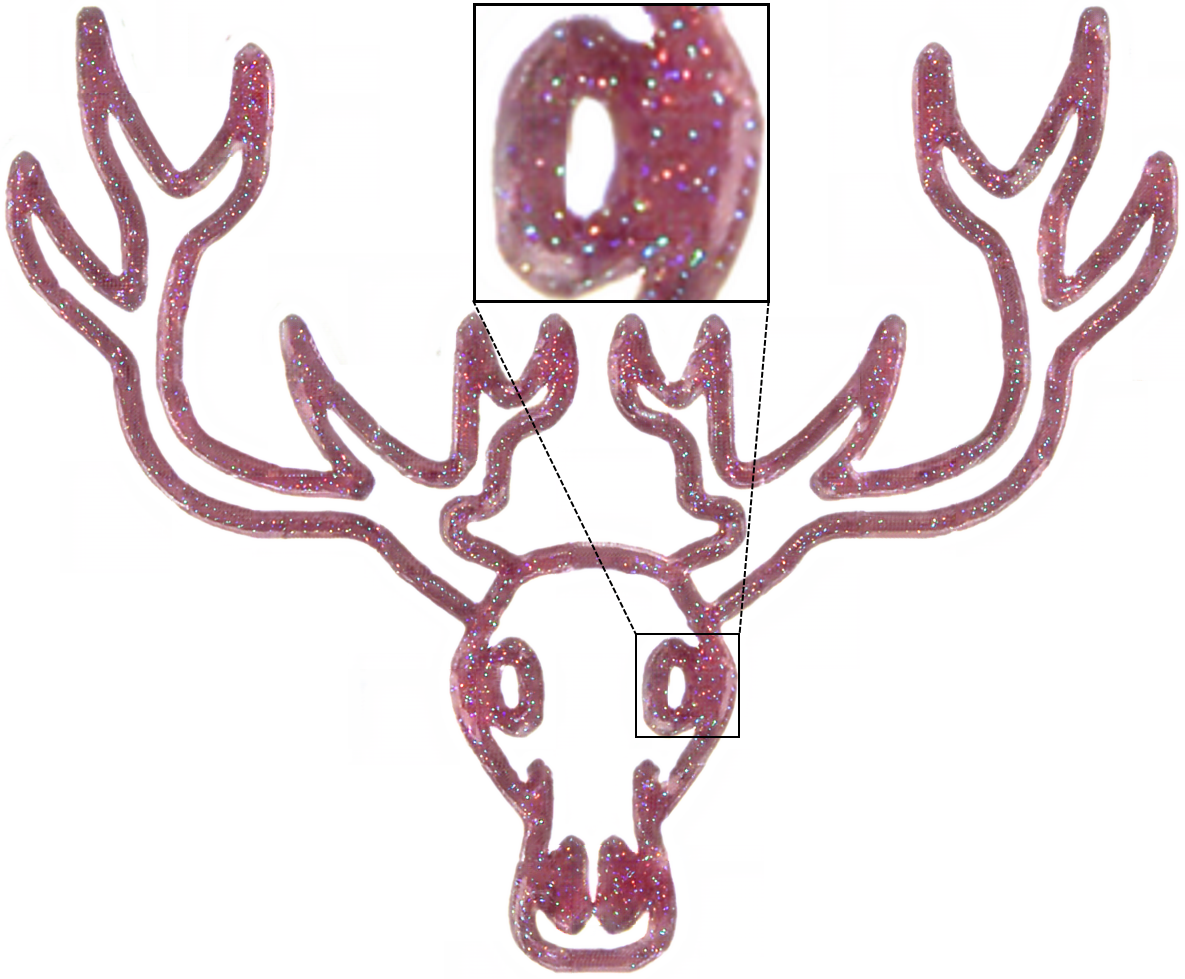}} \\
\multicolumn{2}{c}{(a)} &
\hspace{8pt} &
\multicolumn{2}{c}{(b)} \\

\end{tabular}

\caption{Natural media stylization. We train our system on natural media brush strokes from RealBrush \cite{lu2013realbrush}. In (a), we apply water color stylization to a sketch of a shrimp, and in (b), a glittery lip gloss is applied to style a moose sketch.
\textit{Sketches by \href{https://www.idangilboa.com/}{Idan Gilboa}.}}
\label{fig:realbrush}
\end{figure}

We apply our approach for natural media stylization, with data collected by RealBrush \cite{lu2013realbrush}. This data contains sets of natural media strokes (see exemplars in Figure \ref{fig:realbrush}, top), with corresponding masks. We used the masks as our source for constructing the plain domain, and since these masks are aligned with the strokes themselves, the pairing between the two is known thus we activate our paired pipeline. 
Figure \ref{fig:realbrush} presents two results obtained by training our system on water color (a) and glittery lip gloss (b) stylization.

\subsubsection{Stroke weight}
\begin{figure}
\newcommand{\ttrfig}{3.7}
\newcommand{\ttsfig}{4.1}
\newcommand{\cheshire}{1.6}
\newcommand{\hands}{3.2}
\setlength\tabcolsep{1pt}

\centering
\begin{tabular}[t]{c c c}

\includegraphics[height=\cheshire cm]{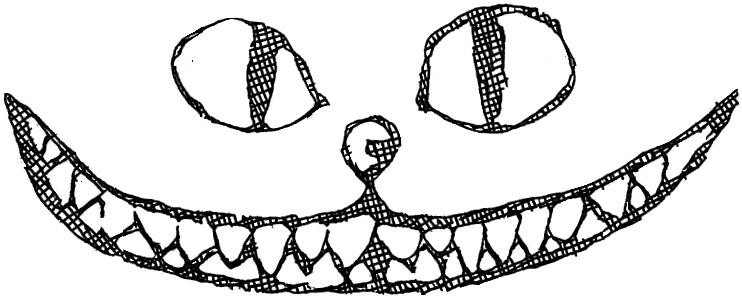} &
\hspace{8pt} &
\includegraphics[height=\cheshire cm]{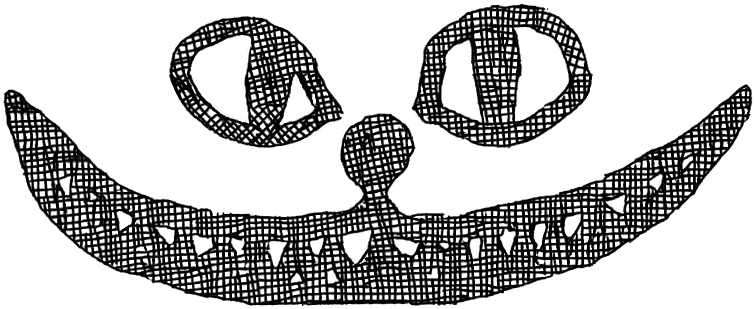} \\
\vspace{8pt}
(a) & \hspace{8pt} & (b) \\

\includegraphics[height=\hands cm]{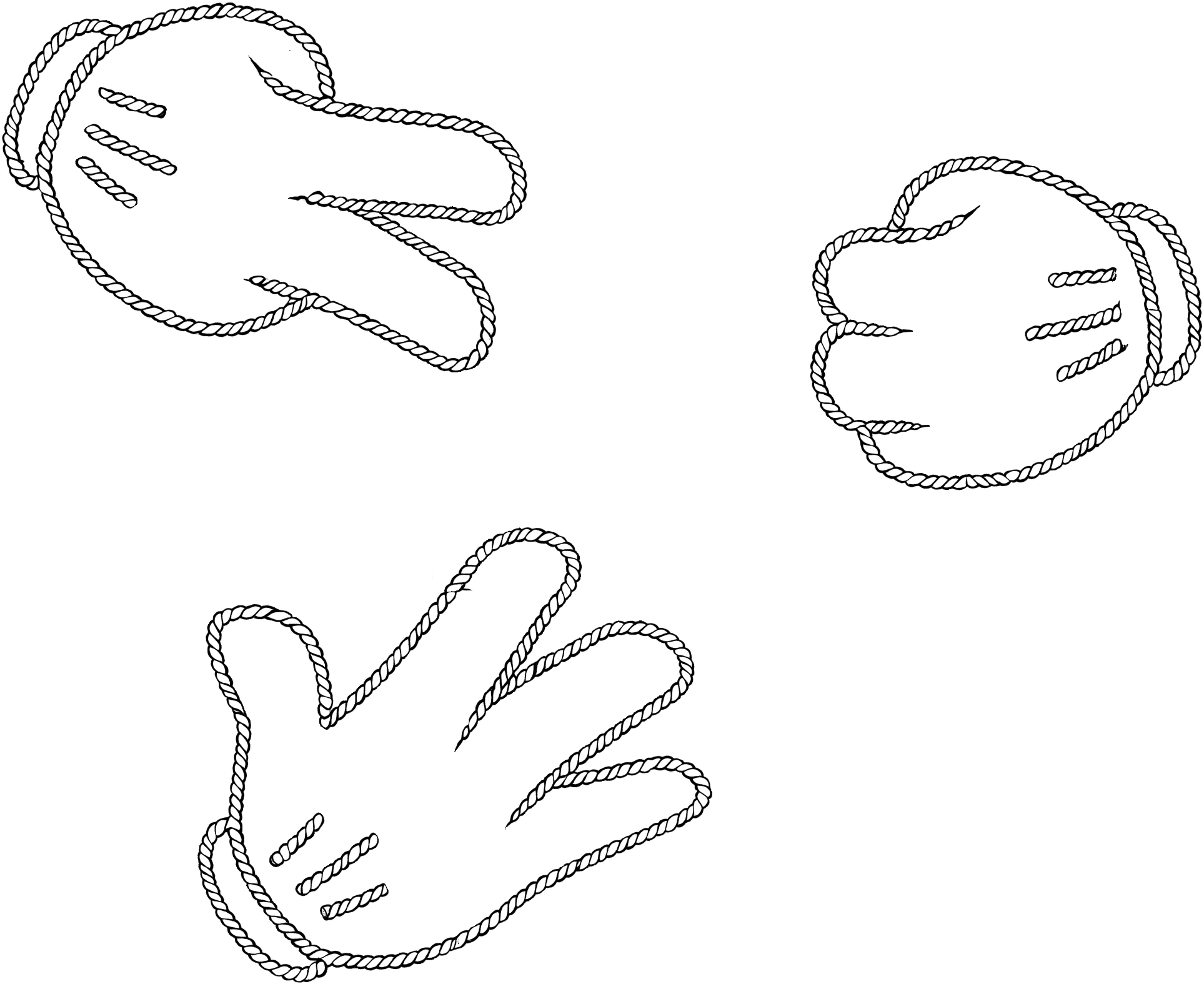}  &
\hspace{8pt} &
\includegraphics[height=\hands cm]{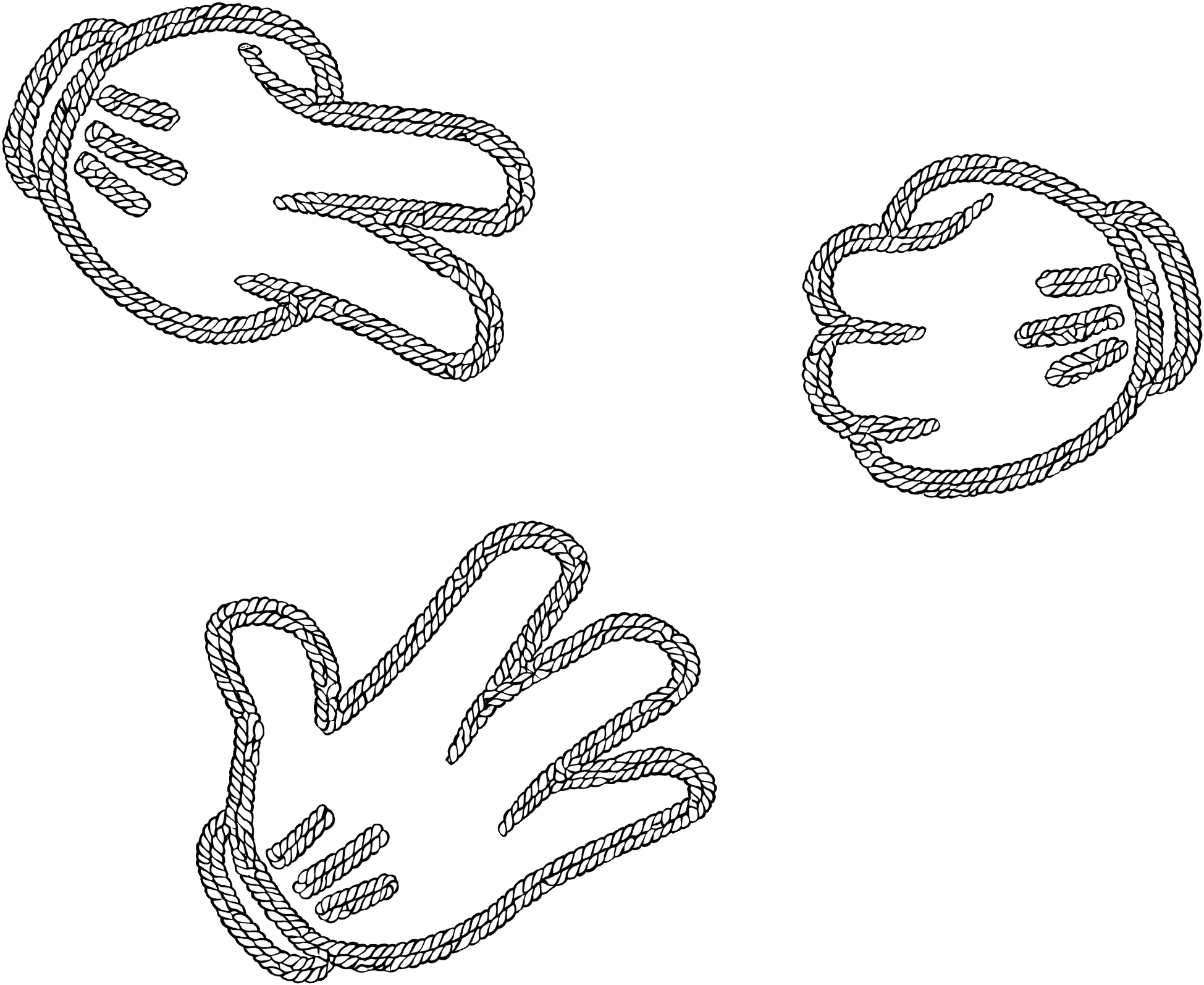} \\
\vspace{8pt}
(c) & \hspace{8pt} & (d) \\

\includegraphics[height=\ttsfig cm]{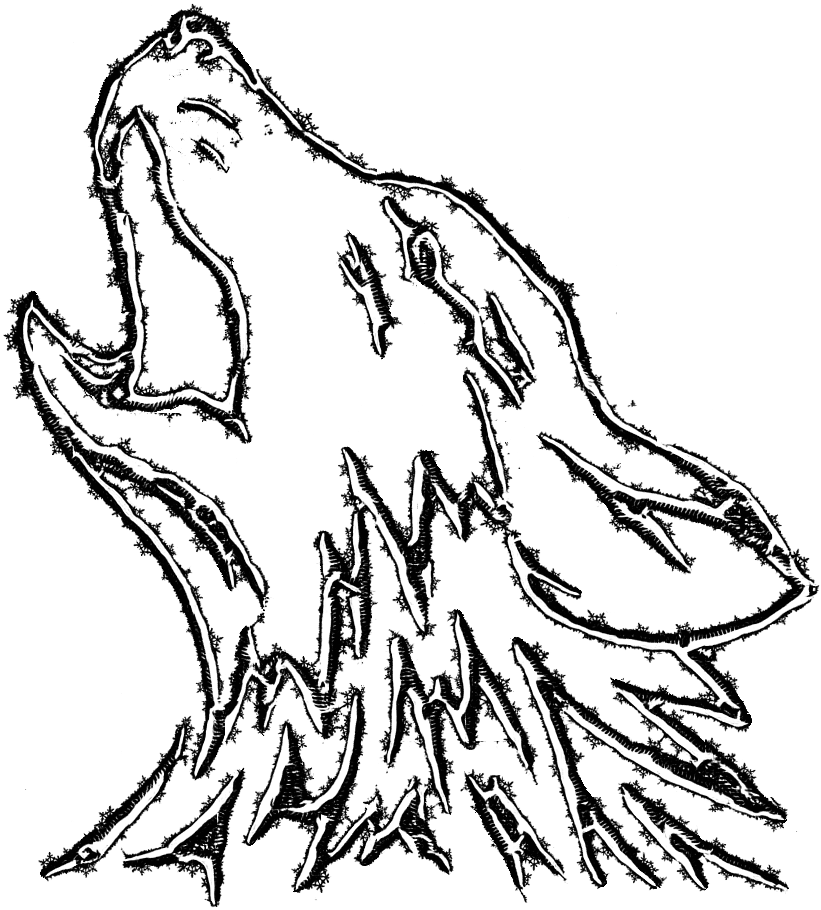} &
\hspace{8pt} &
\includegraphics[height=\ttsfig cm]{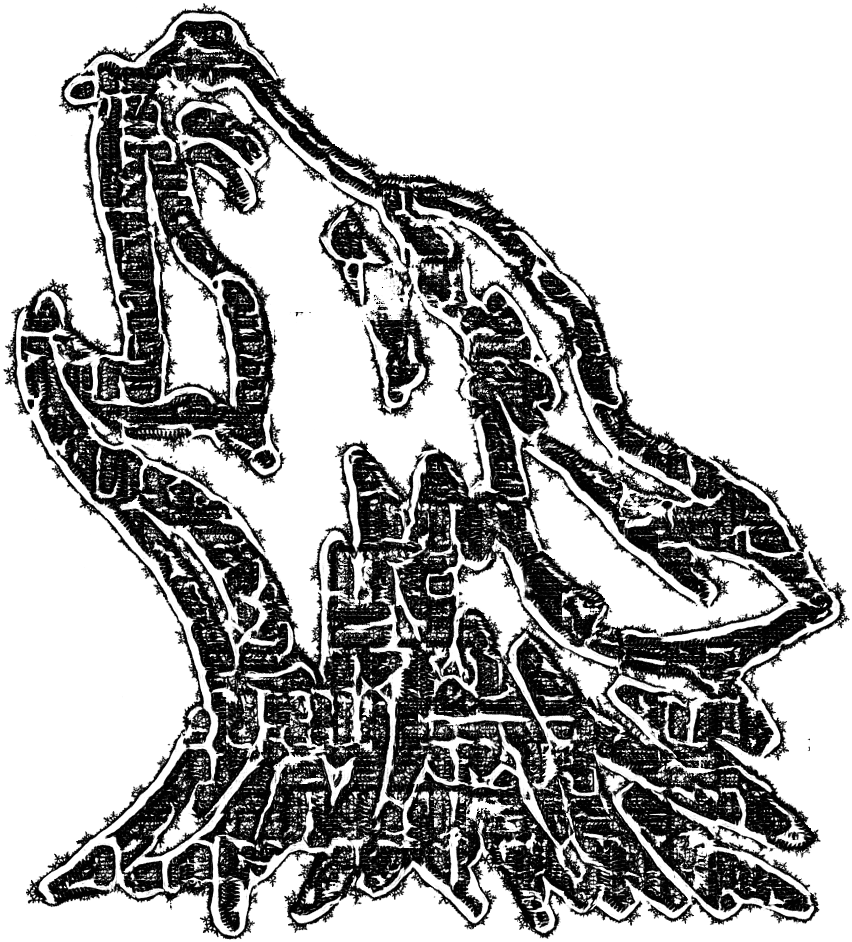} \\
(e) & \hspace{8pt} & (f) \\

\end{tabular}
\caption{Stroke weight robustness. We show two styles that are robust to changes in stroke weight and are able to adjust accordingly (a-b and c-d), vs. a style that is limited to thin strokes, and produces visible artifacts when applied to thick ones (e-f).
\textit{
Cheshire cat smile and howling wolf sketches by \href{https://www.idangilboa.com/}{Idan Gilboa}.}
}

\label{fig:stroke_weight}
\end{figure}

As mentioned above, in our experience, stroke weight compatibility often proves to be a key ingredient in support of a successful CycleGAN learning process. Similarly, at inference time, depending upon the certain style mastered by our network, incompatible sketch stroke weight may compromise the expected result. However, since we currently target binary, black and white test sketches, we can easily preprocess our sketches to adjust stroke weight, using simple operations of erosion and dilation, thereby extending the range of applicability of our method. While some styles struggle to generalize outside their inherent underlying stroke weight, others generalize beautifully, allowing us greater freedom to design our sketches as we see fit (see Figure \ref{fig:stroke_weight}).

\begin{figure*}[h]
\newcommand{\plfig}{14}

\centering
\includegraphics[height=\plfig cm]{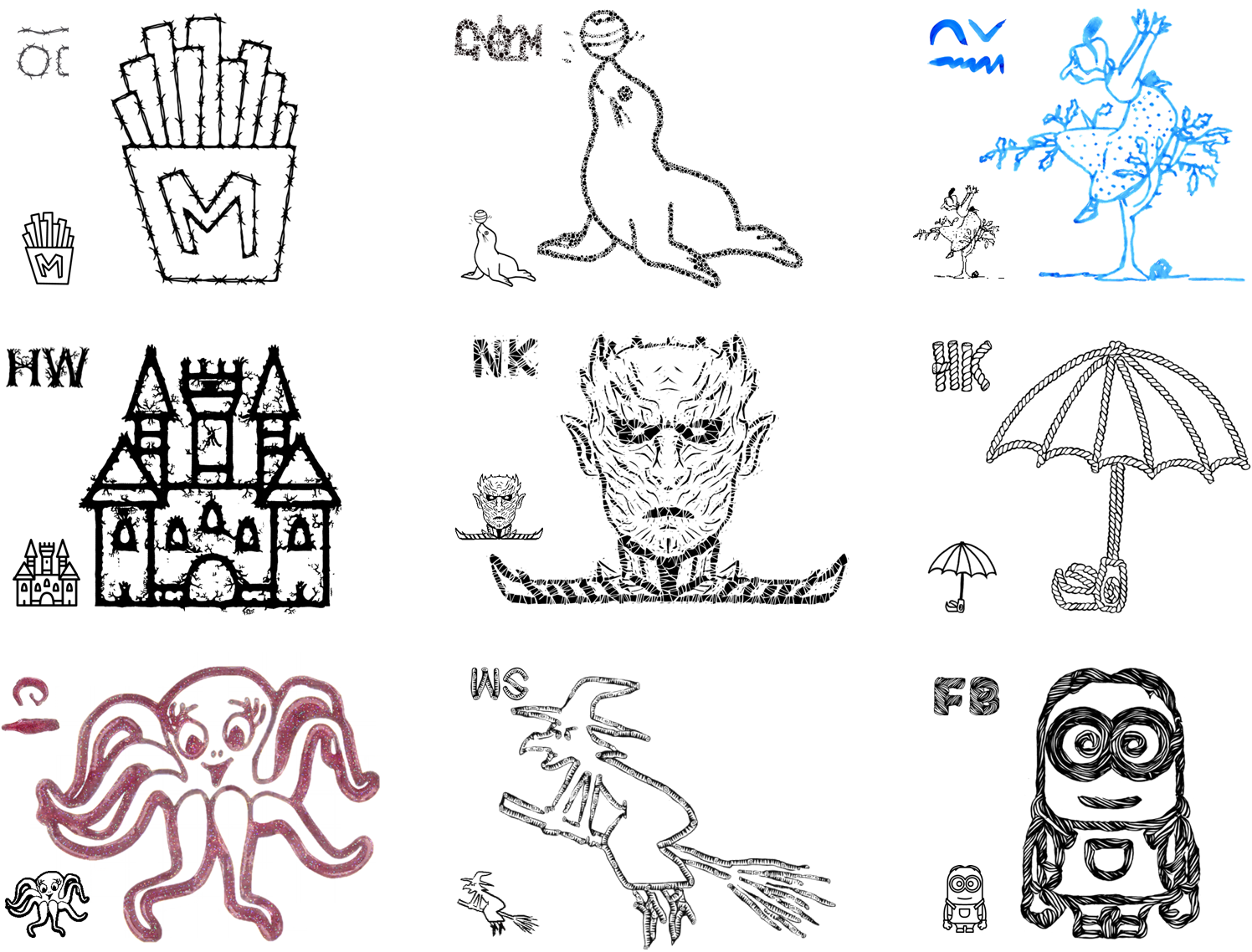}
\caption{A selection of sketches styled using our approach. We present nine sketch and style pairs featuring diverse geometries and stylistic elements. 
A sample of the corresponding style exemplar appears at the top-left of each finished result, and the original plain sketch can be found at its bottom-left.
\textit{
Fries, night king, umbrella and witch sketches by \href{https://www.idangilboa.com/}{Idan Gilboa},
vintage sketch from the Book of Limericks 1888,
castle icon by \href{https://www.iconfinder.com/}{Iconfinder}, Minion icon by \href{https://www.pngrepo.com/}{PNGRepo}.
Barbed wire style exemplar created by \href{https://www.freepik.com/vectors/line}{Line vector on Freepik}.
}
}
\label{fig:charachter_parade}
\end{figure*}
\begin{figure*}[h]
\newcommand{\plfig}{18}

\centering
\includegraphics[width=\plfig cm]{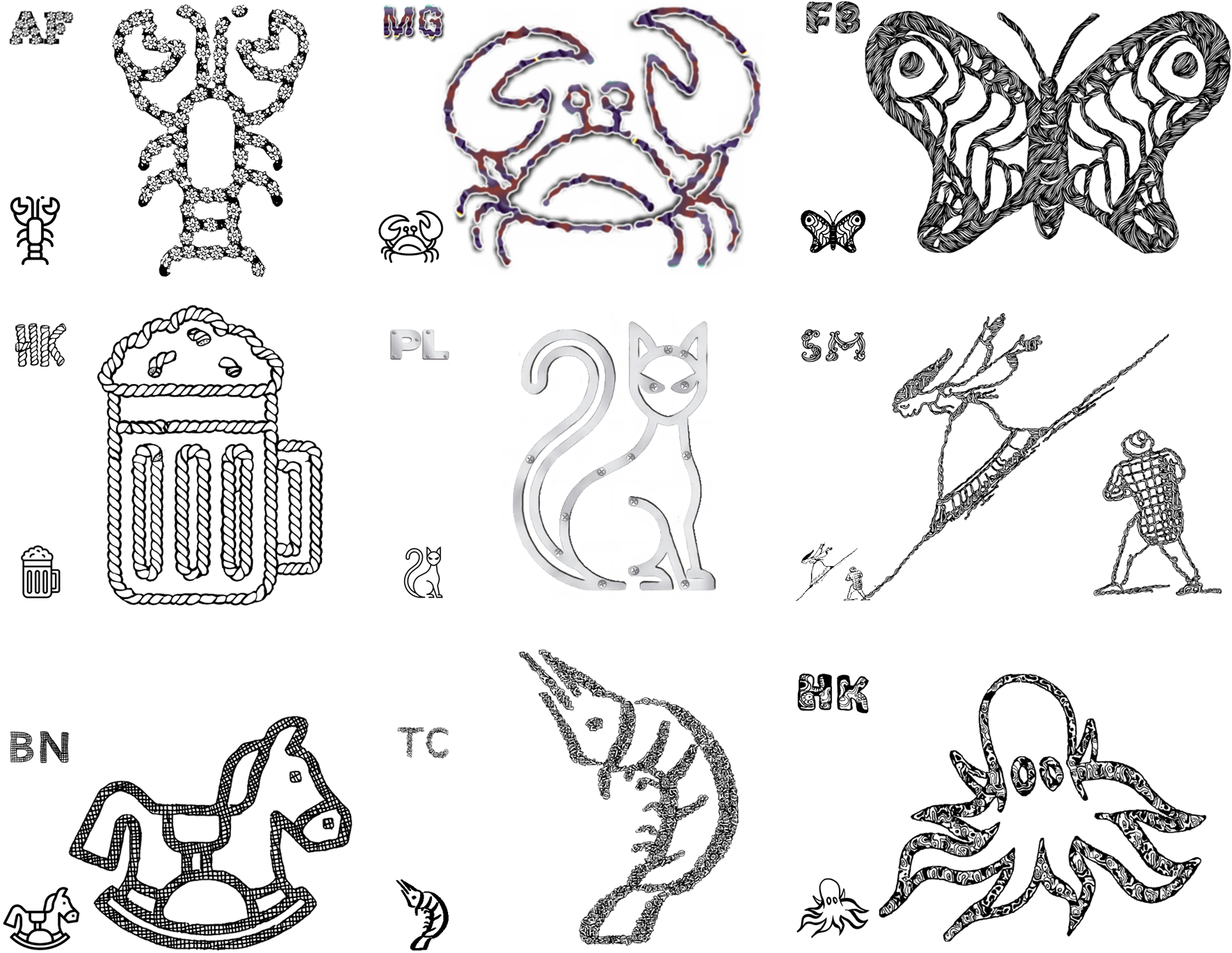}
\caption{A selection of sketches styled using our approach. We present nine sketch and style pairs featuring diverse geometries and stylistic elements.
A sample of the corresponding style exemplar appears at the top-left of each finished result, and the original plain sketch can be found at its bottom-left.
\textit{
Lobster icon by \href{https://www.pngrepo.com/}{PNGRepo},
crab, butterfly, shrimp and octopus sketches by \href{https://www.idangilboa.com/}{Idan Gilboa}, beer and cat icons by \href{https://www.iconfinder.com/}{Iconfinder},
vintage sketch from the Book of Limericks 1888, rocking horse icon by \href{https://www.flaticon.com/authors/freepik}{Freepik}. Typography exemplar in the center by \href{https://www.freepik.com/vectors/typography}{Alvaro Cabrera via Freepik}.
}
}
\label{fig:naive_parade}
\end{figure*}

\section{Discussion and Future work}

Motivated by impressive advancements in neural style transfer for images and drawings,
in this work, we target sketch stylization.
Lack of sufficient data to learn specific geometric patterns from, renders standard image-to-image translation approaches intractable at this time, thus we observed that by operating at the patch level, stylistic elements can be suitably gleaned from a handful of appropriately styled sketches, offering an affluence of basic geometric primitives.
Working with patches, however, raised the question of arbitrarily sized sketch translation, and the inconsistencies and visible seams running along patch borders, resulting from independent patch-by-patch translation.
To address that, we presented a seamless translation model, that operates on paired style and plain patches, or, more specifically, hybrid patches created by combining styled and plain elements from each pair. This model is trained to output a fully styled patch, and is therefore prepared to tackle full-sized sketches at inference time, by simply dividing a given sketch into overlapping patches for translation. 

When a pairing between a given style exemplar and a plain counterpart is unavailable, we enlist the help of CycleGAN to learn the translation between patches cut up from the styled exemplar, and those cut up from plain sketches featuring similar geometries.

With a wide selection of sketches and styles, we demonstrated that our approach is robust to arbitrarily sized sketches featuring diverse geometries, and produces visually pleasing styled sketches, inspired by, but not equal to, the given styled exemplars, with little to no discernible seams at patch borders. We compared our approach to neural style transfer methods, and to our baseline approaches based on CycleGAN translation, and have established that styles that feature continuous geometric elements are particularly impaired by the lack of dependency between neighboring patches.

Despite successful stylization demonstrated by our technique, we note a few shortcomings of its current design.

As we discussed, stroke weight is an aspect that occasionally, depending upon the style in question, limits the applicability of our trained system. Specifically, inference time sketches featuring strokes of different weight to that of the target style  
may not undergo translation successfully, due to incompatibility with the training distribution.
Currently, we address this issue by preprocessing our inference sketches to generate thinner and thicker versions via erosion and dilation,
but perhaps a more elegant solution will leverage multi-class concepts for a one-to-many translation, or even many-to-many, with different styles co-existing within one network.

Additionally, styles that feature inconsistent ornamentation pose a problem, since they inevitably confuse the model throughout the learning process, as it tries to reason between the contradicting evidence it is given. Multi-scale translation or context additions may be a viable solution to this, opening up interesting opportunities for future work.
Furthermore, we revisit the potential loss of global orientation of patterns resulting from training on rotation-augmented data, with a possible "quick fix". Taking the initial patch to be translated at test time, we can essentially transform it into a hybrid by pasting a sub-patch region from the real exemplar, within an otherwise empty space in the patch. 
Supported by our patch translation order, the network will continue to translate along the same general lines of orientation, after which the artificially placed sub-patch content can be removed.

In the unpaired translation scenario, we note that
any inaccuracies or impediments in the learning process of CycleGAN directly impact the training of ST by way of a noisy supervision. As such, due to a known CycleGAN rigidity with respect to translation of shape, we experience difficulties with translation that is more geometric in nature, e.g., adding curvature. 

Finally, as demonstrated throughout our results, at this time we target clean-lined sketches, lacking various types of sketching techniques such as shading and filling. This is due to the nature of the geometric style exemplars - featuring patterned designs along ultimately clean lines and curves. Translating between such a styled domain and a noisier plain domain therefore requires further handling, and is an important venture for future work.

\section*{Acknowledgements}
We are grateful to Idan Gilboa (\url{https://www.idangilboa.com}) for creating and sharing many of the art pieces displayed in this paper, and to the anonymous reviewers for their helpful feedback.
This work was supported by the Israel Science Foundation (grant no. 2366/16 and 2472/17).

\bibliographystyle{ACM-Reference-Format}
\bibliography{bibs}


\begin{thebibliography}{00}


\ifx \showCODEN    \undefined \def \showCODEN     #1{\unskip}     \fi
\ifx \showDOI      \undefined \def \showDOI       #1{#1}\fi
\ifx \showISBNx    \undefined \def \showISBNx     #1{\unskip}     \fi
\ifx \showISBNxiii \undefined \def \showISBNxiii  #1{\unskip}     \fi
\ifx \showISSN     \undefined \def \showISSN      #1{\unskip}     \fi
\ifx \showLCCN     \undefined \def \showLCCN      #1{\unskip}     \fi
\ifx \shownote     \undefined \def \shownote      #1{#1}          \fi
\ifx \showarticletitle \undefined \def \showarticletitle #1{#1}   \fi
\ifx \showURL      \undefined \def \showURL       {\relax}        \fi
\providecommand\bibfield[2]{#2}
\providecommand\bibinfo[2]{#2}
\providecommand\natexlab[1]{#1}
\providecommand\showeprint[2][]{arXiv:#2}

\bibitem[\protect\citeauthoryear{Azadi, Fisher, Kim, Wang, Shechtman, and
  Darrell}{Azadi et~al\mbox{.}}{2018}]%
        {azadi2018multi}
\bibfield{author}{\bibinfo{person}{Samaneh Azadi}, \bibinfo{person}{Matthew
  Fisher}, \bibinfo{person}{Vladimir Kim}, \bibinfo{person}{Zhaowen Wang},
  \bibinfo{person}{Eli Shechtman}, {and} \bibinfo{person}{Trevor Darrell}.}
  \bibinfo{year}{2018}\natexlab{}.
\newblock \showarticletitle{Multi-content gan for few-shot font style
  transfer}. In \bibinfo{booktitle}{{\em Proceedings of the IEEE Conference on
  Computer Vision and Pattern Recognition}}, Vol.~\bibinfo{volume}{11}.
  \bibinfo{pages}{13}.
\newblock


\bibitem[\protect\citeauthoryear{Barnes, Shechtman, Goldman, and
  Finkelstein}{Barnes et~al\mbox{.}}{2010}]%
        {barnes2010generalized}
\bibfield{author}{\bibinfo{person}{Connelly Barnes}, \bibinfo{person}{Eli
  Shechtman}, \bibinfo{person}{Dan~B Goldman}, {and} \bibinfo{person}{Adam
  Finkelstein}.} \bibinfo{year}{2010}\natexlab{}.
\newblock \showarticletitle{The generalized patchmatch correspondence
  algorithm}. In \bibinfo{booktitle}{{\em European Conference on Computer
  Vision}}. Springer, \bibinfo{pages}{29--43}.
\newblock


\bibitem[\protect\citeauthoryear{Barnes and Zhang}{Barnes and Zhang}{2017}]%
        {barnes2017survey}
\bibfield{author}{\bibinfo{person}{Connelly Barnes} {and}
  \bibinfo{person}{Fang-Lue Zhang}.} \bibinfo{year}{2017}\natexlab{}.
\newblock \showarticletitle{A survey of the state-of-the-art in patch-based
  synthesis}.
\newblock \bibinfo{journal}{{\em Computational Visual Media\/}}
  \bibinfo{volume}{3}, \bibinfo{number}{1} (\bibinfo{year}{2017}),
  \bibinfo{pages}{3--20}.
\newblock


\bibitem[\protect\citeauthoryear{Chen and Schmidt}{Chen and Schmidt}{2016}]%
        {chen2016fast}
\bibfield{author}{\bibinfo{person}{Tian~Qi Chen} {and} \bibinfo{person}{Mark
  Schmidt}.} \bibinfo{year}{2016}\natexlab{}.
\newblock \showarticletitle{Fast patch-based style transfer of arbitrary
  style}.
\newblock \bibinfo{journal}{{\em arXiv preprint arXiv:1612.04337\/}}
  (\bibinfo{year}{2016}).
\newblock


\bibitem[\protect\citeauthoryear{Gatys, Ecker, and Bethge}{Gatys
  et~al\mbox{.}}{2016}]%
        {gatys2016image}
\bibfield{author}{\bibinfo{person}{Leon~A Gatys}, \bibinfo{person}{Alexander~S
  Ecker}, {and} \bibinfo{person}{Matthias Bethge}.}
  \bibinfo{year}{2016}\natexlab{}.
\newblock \showarticletitle{Image style transfer using convolutional neural
  networks}. In \bibinfo{booktitle}{{\em Proceedings of the IEEE conference on
  computer vision and pattern recognition}}. \bibinfo{pages}{2414--2423}.
\newblock


\bibitem[\protect\citeauthoryear{Hertzmann, Jacobs, Oliver, Curless, and
  Salesin}{Hertzmann et~al\mbox{.}}{2001}]%
        {hertzmann2001image}
\bibfield{author}{\bibinfo{person}{Aaron Hertzmann}, \bibinfo{person}{Charles~E
  Jacobs}, \bibinfo{person}{Nuria Oliver}, \bibinfo{person}{Brian Curless},
  {and} \bibinfo{person}{David~H Salesin}.} \bibinfo{year}{2001}\natexlab{}.
\newblock \showarticletitle{Image analogies}. In \bibinfo{booktitle}{{\em
  Proceedings of the 28th annual conference on Computer graphics and
  interactive techniques}}. \bibinfo{pages}{327--340}.
\newblock


\bibitem[\protect\citeauthoryear{Huang and Belongie}{Huang and
  Belongie}{2017}]%
        {huang2017arbitrary}
\bibfield{author}{\bibinfo{person}{Xun Huang} {and} \bibinfo{person}{Serge
  Belongie}.} \bibinfo{year}{2017}\natexlab{}.
\newblock \showarticletitle{Arbitrary style transfer in real-time with adaptive
  instance normalization}. In \bibinfo{booktitle}{{\em Proceedings of the IEEE
  International Conference on Computer Vision}}. \bibinfo{pages}{1501--1510}.
\newblock


\bibitem[\protect\citeauthoryear{Huang, Liu, Belongie, and Kautz}{Huang
  et~al\mbox{.}}{2018}]%
        {huang2018multimodal}
\bibfield{author}{\bibinfo{person}{Xun Huang}, \bibinfo{person}{Ming-Yu Liu},
  \bibinfo{person}{Serge Belongie}, {and} \bibinfo{person}{Jan Kautz}.}
  \bibinfo{year}{2018}\natexlab{}.
\newblock \showarticletitle{Multimodal unsupervised image-to-image
  translation}. In \bibinfo{booktitle}{{\em Proceedings of the European
  Conference on Computer Vision (ECCV)}}. \bibinfo{pages}{172--189}.
\newblock


\bibitem[\protect\citeauthoryear{Isola, Zhu, Zhou, and Efros}{Isola
  et~al\mbox{.}}{2017}]%
        {isola2017image}
\bibfield{author}{\bibinfo{person}{Phillip Isola}, \bibinfo{person}{Jun-Yan
  Zhu}, \bibinfo{person}{Tinghui Zhou}, {and} \bibinfo{person}{Alexei~A
  Efros}.} \bibinfo{year}{2017}\natexlab{}.
\newblock \showarticletitle{Image-to-image translation with conditional
  adversarial networks}. In \bibinfo{booktitle}{{\em Proceedings of the IEEE
  conference on computer vision and pattern recognition}}.
  \bibinfo{pages}{1125--1134}.
\newblock


\bibitem[\protect\citeauthoryear{Johnson, Alahi, and Fei-Fei}{Johnson
  et~al\mbox{.}}{2016}]%
        {johnson2016perceptual}
\bibfield{author}{\bibinfo{person}{Justin Johnson}, \bibinfo{person}{Alexandre
  Alahi}, {and} \bibinfo{person}{Li Fei-Fei}.} \bibinfo{year}{2016}\natexlab{}.
\newblock \showarticletitle{Perceptual losses for real-time style transfer and
  super-resolution}. In \bibinfo{booktitle}{{\em European conference on
  computer vision}}. Springer, \bibinfo{pages}{694--711}.
\newblock


\bibitem[\protect\citeauthoryear{Kazi, Igarashi, Zhao, and Davis}{Kazi
  et~al\mbox{.}}{2012}]%
        {kazi2012vignette}
\bibfield{author}{\bibinfo{person}{Rubaiat~Habib Kazi}, \bibinfo{person}{Takeo
  Igarashi}, \bibinfo{person}{Shengdong Zhao}, {and} \bibinfo{person}{Richard
  Davis}.} \bibinfo{year}{2012}\natexlab{}.
\newblock \showarticletitle{Vignette: interactive texture design and
  manipulation with freeform gestures for pen-and-ink illustration}. In
  \bibinfo{booktitle}{{\em Proceedings of the SIGCHI Conference on Human
  Factors in Computing Systems}}. ACM, \bibinfo{pages}{1727--1736}.
\newblock


\bibitem[\protect\citeauthoryear{Kim, Cha, Kim, Lee, and Kim}{Kim
  et~al\mbox{.}}{2017}]%
        {kim2017learning}
\bibfield{author}{\bibinfo{person}{Taeksoo Kim}, \bibinfo{person}{Moonsu Cha},
  \bibinfo{person}{Hyunsoo Kim}, \bibinfo{person}{Jung~Kwon Lee}, {and}
  \bibinfo{person}{Jiwon Kim}.} \bibinfo{year}{2017}\natexlab{}.
\newblock \showarticletitle{Learning to discover cross-domain relations with
  generative adversarial networks}. In \bibinfo{booktitle}{{\em Proceedings of
  the 34th International Conference on Machine Learning-Volume 70}}. JMLR. org,
  \bibinfo{pages}{1857--1865}.
\newblock


\bibitem[\protect\citeauthoryear{Lang and Alexa}{Lang and Alexa}{2015}]%
        {lang2015markov}
\bibfield{author}{\bibinfo{person}{Katrin Lang} {and} \bibinfo{person}{Marc
  Alexa}.} \bibinfo{year}{2015}\natexlab{}.
\newblock \showarticletitle{The Markov pen: online synthesis of free-hand
  drawing styles}. In \bibinfo{booktitle}{{\em Proceedings of the workshop on
  Non-Photorealistic Animation and Rendering}}. Eurographics Association,
  \bibinfo{pages}{203--215}.
\newblock


\bibitem[\protect\citeauthoryear{Li, Fang, Hertzmann, Shechtman, and Yang}{Li
  et~al\mbox{.}}{2019}]%
        {li2019im2pencil}
\bibfield{author}{\bibinfo{person}{Yijun Li}, \bibinfo{person}{Chen Fang},
  \bibinfo{person}{Aaron Hertzmann}, \bibinfo{person}{Eli Shechtman}, {and}
  \bibinfo{person}{Ming-Hsuan Yang}.} \bibinfo{year}{2019}\natexlab{}.
\newblock \showarticletitle{Im2pencil: Controllable pencil illustration from
  photographs}. In \bibinfo{booktitle}{{\em Proceedings of the IEEE Conference
  on Computer Vision and Pattern Recognition}}. \bibinfo{pages}{1525--1534}.
\newblock


\bibitem[\protect\citeauthoryear{Li, Fang, Yang, Wang, Lu, and Yang}{Li
  et~al\mbox{.}}{2017}]%
        {li2017universal}
\bibfield{author}{\bibinfo{person}{Yijun Li}, \bibinfo{person}{Chen Fang},
  \bibinfo{person}{Jimei Yang}, \bibinfo{person}{Zhaowen Wang},
  \bibinfo{person}{Xin Lu}, {and} \bibinfo{person}{Ming-Hsuan Yang}.}
  \bibinfo{year}{2017}\natexlab{}.
\newblock \showarticletitle{Universal style transfer via feature transforms}.
  In \bibinfo{booktitle}{{\em Advances in neural information processing
  systems}}. \bibinfo{pages}{386--396}.
\newblock


\bibitem[\protect\citeauthoryear{Liao, Yao, Yuan, Hua, and Kang}{Liao
  et~al\mbox{.}}{2017}]%
        {liao2017visual}
\bibfield{author}{\bibinfo{person}{Jing Liao}, \bibinfo{person}{Yuan Yao},
  \bibinfo{person}{Lu Yuan}, \bibinfo{person}{Gang Hua}, {and}
  \bibinfo{person}{Sing~Bing Kang}.} \bibinfo{year}{2017}\natexlab{}.
\newblock \showarticletitle{Visual attribute transfer through deep image
  analogy}.
\newblock \bibinfo{journal}{{\em ACM Transactions on Graphics (TOG)\/}}
  \bibinfo{volume}{36}, \bibinfo{number}{4} (\bibinfo{year}{2017}),
  \bibinfo{pages}{120}.
\newblock


\bibitem[\protect\citeauthoryear{Lu, Barnes, DiVerdi, and Finkelstein}{Lu
  et~al\mbox{.}}{2013}]%
        {lu2013realbrush}
\bibfield{author}{\bibinfo{person}{Jingwan Lu}, \bibinfo{person}{Connelly
  Barnes}, \bibinfo{person}{Stephen DiVerdi}, {and} \bibinfo{person}{Adam
  Finkelstein}.} \bibinfo{year}{2013}\natexlab{}.
\newblock \showarticletitle{RealBrush: painting with examples of physical
  media}.
\newblock \bibinfo{journal}{{\em ACM Transactions on Graphics (TOG)\/}}
  \bibinfo{volume}{32}, \bibinfo{number}{4} (\bibinfo{year}{2013}),
  \bibinfo{pages}{117}.
\newblock


\bibitem[\protect\citeauthoryear{Lu, Barnes, Wan, Asente, Mech, and
  Finkelstein}{Lu et~al\mbox{.}}{2014}]%
        {lu2014decobrush}
\bibfield{author}{\bibinfo{person}{Jingwan Lu}, \bibinfo{person}{Connelly
  Barnes}, \bibinfo{person}{Connie Wan}, \bibinfo{person}{Paul Asente},
  \bibinfo{person}{Radomir Mech}, {and} \bibinfo{person}{Adam Finkelstein}.}
  \bibinfo{year}{2014}\natexlab{}.
\newblock \showarticletitle{DecoBrush: drawing structured decorative patterns
  by example}.
\newblock \bibinfo{journal}{{\em ACM Transactions on Graphics (TOG)\/}}
  \bibinfo{volume}{33}, \bibinfo{number}{4} (\bibinfo{year}{2014}),
  \bibinfo{pages}{90}.
\newblock


\bibitem[\protect\citeauthoryear{Luk{\'a}{\v{c}}, Fi{\v{s}}er, Asente, Lu,
  Shechtman, and S{\`y}kora}{Luk{\'a}{\v{c}} et~al\mbox{.}}{2015}]%
        {lukavc2015brushables}
\bibfield{author}{\bibinfo{person}{Michal Luk{\'a}{\v{c}}},
  \bibinfo{person}{Jakub Fi{\v{s}}er}, \bibinfo{person}{Paul Asente},
  \bibinfo{person}{Jingwan Lu}, \bibinfo{person}{Eli Shechtman}, {and}
  \bibinfo{person}{Daniel S{\`y}kora}.} \bibinfo{year}{2015}\natexlab{}.
\newblock \showarticletitle{Brushables: Example-based Edge-aware Directional
  Texture Painting}. In \bibinfo{booktitle}{{\em Computer Graphics Forum}},
  Vol.~\bibinfo{volume}{34}. Wiley Online Library, \bibinfo{pages}{257--267}.
\newblock


\bibitem[\protect\citeauthoryear{Mao, Li, Xie, Lau, Wang, and Paul~Smolley}{Mao
  et~al\mbox{.}}{2017}]%
        {mao2017least}
\bibfield{author}{\bibinfo{person}{Xudong Mao}, \bibinfo{person}{Qing Li},
  \bibinfo{person}{Haoran Xie}, \bibinfo{person}{Raymond~YK Lau},
  \bibinfo{person}{Zhen Wang}, {and} \bibinfo{person}{Stephen Paul~Smolley}.}
  \bibinfo{year}{2017}\natexlab{}.
\newblock \showarticletitle{Least squares generative adversarial networks}. In
  \bibinfo{booktitle}{{\em Proceedings of the IEEE International Conference on
  Computer Vision}}. \bibinfo{pages}{2794--2802}.
\newblock


\bibitem[\protect\citeauthoryear{Phan, Lu, Asente, Chan, and Fu}{Phan
  et~al\mbox{.}}{2016}]%
        {10.5555/2981324.2981336}
\bibfield{author}{\bibinfo{person}{H.~Q. Phan}, \bibinfo{person}{J. Lu},
  \bibinfo{person}{P. Asente}, \bibinfo{person}{A.~B. Chan}, {and}
  \bibinfo{person}{H. Fu}.} \bibinfo{year}{2016}\natexlab{}.
\newblock \showarticletitle{Patternista: Learning Element Style Compatibility
  and Spatial Composition for Ring-Based Layout Decoration}. In
  \bibinfo{booktitle}{{\em Proceedings of the Joint Symposium on Computational
  Aesthetics and Sketch Based Interfaces and Modeling and Non-Photorealistic
  Animation and Rendering}} {\em (\bibinfo{series}{Expresive ’16})}.
  \bibinfo{publisher}{Eurographics Association}, \bibinfo{address}{Goslar,
  DEU}, \bibinfo{pages}{79–88}.
\newblock


\bibitem[\protect\citeauthoryear{Santoni and Pellacini}{Santoni and
  Pellacini}{2016}]%
        {santoni2016gtangle}
\bibfield{author}{\bibinfo{person}{Christian Santoni} {and}
  \bibinfo{person}{Fabio Pellacini}.} \bibinfo{year}{2016}\natexlab{}.
\newblock \showarticletitle{gTangle: A grammar for the procedural generation of
  tangle patterns}.
\newblock \bibinfo{journal}{{\em ACM Transactions on Graphics (TOG)\/}}
  \bibinfo{volume}{35}, \bibinfo{number}{6} (\bibinfo{year}{2016}),
  \bibinfo{pages}{1--11}.
\newblock


\bibitem[\protect\citeauthoryear{Simonyan and Zisserman}{Simonyan and
  Zisserman}{2014}]%
        {simonyan2014very}
\bibfield{author}{\bibinfo{person}{Karen Simonyan} {and}
  \bibinfo{person}{Andrew Zisserman}.} \bibinfo{year}{2014}\natexlab{}.
\newblock \showarticletitle{Very deep convolutional networks for large-scale
  image recognition}.
\newblock \bibinfo{journal}{{\em arXiv preprint arXiv:1409.1556\/}}
  (\bibinfo{year}{2014}).
\newblock


\bibitem[\protect\citeauthoryear{Texler, Fi{\v{s}}er, Luk{\'a}{\v{c}}, Lu,
  Shechtman, and S{\`y}kora}{Texler et~al\mbox{.}}{2019}]%
        {texler2019enhancing}
\bibfield{author}{\bibinfo{person}{Ond{\v{r}}ej Texler}, \bibinfo{person}{Jakub
  Fi{\v{s}}er}, \bibinfo{person}{Mike Luk{\'a}{\v{c}}},
  \bibinfo{person}{Jingwan Lu}, \bibinfo{person}{Eli Shechtman}, {and}
  \bibinfo{person}{Daniel S{\`y}kora}.} \bibinfo{year}{2019}\natexlab{}.
\newblock \showarticletitle{Enhancing neural style transfer using patch-based
  synthesis}. In \bibinfo{booktitle}{{\em Proceedings of the 8th
  ACM/Eurographics Expressive Symposium on Computational Aesthetics and Sketch
  Based Interfaces and Modeling and Non-Photorealistic Animation and
  Rendering}}. Eurographics Association, \bibinfo{pages}{43--50}.
\newblock


\bibitem[\protect\citeauthoryear{Wu, Chen, Wang, Yang, and Marschner}{Wu
  et~al\mbox{.}}{2018}]%
        {wu2018brush}
\bibfield{author}{\bibinfo{person}{Rundong Wu}, \bibinfo{person}{Zhili Chen},
  \bibinfo{person}{Zhaowen Wang}, \bibinfo{person}{Jimei Yang}, {and}
  \bibinfo{person}{Steve Marschner}.} \bibinfo{year}{2018}\natexlab{}.
\newblock \showarticletitle{Brush stroke synthesis with a generative
  adversarial network driven by physically based simulation}. In
  \bibinfo{booktitle}{{\em Proceedings of the Joint Symposium on Computational
  Aesthetics and Sketch-Based Interfaces and Modeling and Non-Photorealistic
  Animation and Rendering}}. ACM, \bibinfo{pages}{12}.
\newblock


\bibitem[\protect\citeauthoryear{Yang, Liu, Wang, and Guo}{Yang
  et~al\mbox{.}}{2019}]%
        {Yang2019TETGAN}
\bibfield{author}{\bibinfo{person}{Shuai Yang}, \bibinfo{person}{Jiaying Liu},
  \bibinfo{person}{Wenjing Wang}, {and} \bibinfo{person}{Zongming Guo}.}
  \bibinfo{year}{2019}\natexlab{}.
\newblock \showarticletitle{TET-GAN: Text Effects Transfer via Stylization and
  Destylization}. In \bibinfo{booktitle}{{\em AAAI Conference on Artificial
  Intelligence}}.
\newblock


\bibitem[\protect\citeauthoryear{Yi, Zhang, Tan, and Gong}{Yi
  et~al\mbox{.}}{2017}]%
        {yi2017dualgan}
\bibfield{author}{\bibinfo{person}{Zili Yi}, \bibinfo{person}{Hao Zhang},
  \bibinfo{person}{Ping Tan}, {and} \bibinfo{person}{Minglun Gong}.}
  \bibinfo{year}{2017}\natexlab{}.
\newblock \showarticletitle{Dualgan: Unsupervised dual learning for
  image-to-image translation}. In \bibinfo{booktitle}{{\em Proceedings of the
  IEEE international conference on computer vision}}.
  \bibinfo{pages}{2849--2857}.
\newblock


\bibitem[\protect\citeauthoryear{Yue, Yuan, Zhouhui, Yingmin, and Jianguo}{Yue
  et~al\mbox{.}}{2019}]%
        {Gao2019Artistic}
\bibfield{author}{\bibinfo{person}{Gao Yue}, \bibinfo{person}{Guo Yuan},
  \bibinfo{person}{Lian Zhouhui}, \bibinfo{person}{Tang Yingmin}, {and}
  \bibinfo{person}{Xiao Jianguo}.} \bibinfo{year}{2019}\natexlab{}.
\newblock \showarticletitle{Artistic Glyph Image Synthesis via One-Stage
  Few-Shot Learning}.
\newblock \bibinfo{journal}{{\em ACM Trans. Graph.\/}} \bibinfo{volume}{38},
  \bibinfo{number}{6}, Article \bibinfo{articleno}{185} (\bibinfo{year}{2019}),
  \bibinfo{numpages}{12}~pages.
\newblock
\showURL{%
\url{http://doi.acm.org/10.1145/3355089.3356574}}


\bibitem[\protect\citeauthoryear{Zhou, Lasram, and Lefebvre}{Zhou
  et~al\mbox{.}}{2013}]%
        {zhou2013example}
\bibfield{author}{\bibinfo{person}{Shizhe Zhou}, \bibinfo{person}{Anass
  Lasram}, {and} \bibinfo{person}{Sylvain Lefebvre}.}
  \bibinfo{year}{2013}\natexlab{}.
\newblock \showarticletitle{By--example synthesis of curvilinear structured
  patterns}. In \bibinfo{booktitle}{{\em Computer Graphics Forum}},
  Vol.~\bibinfo{volume}{32}. Wiley Online Library, \bibinfo{pages}{355--360}.
\newblock


\bibitem[\protect\citeauthoryear{Zhu, Park, Isola, and Efros}{Zhu
  et~al\mbox{.}}{2017}]%
        {zhu2017unpaired}
\bibfield{author}{\bibinfo{person}{Jun-Yan Zhu}, \bibinfo{person}{Taesung
  Park}, \bibinfo{person}{Phillip Isola}, {and} \bibinfo{person}{Alexei~A
  Efros}.} \bibinfo{year}{2017}\natexlab{}.
\newblock \showarticletitle{Unpaired image-to-image translation using
  cycle-consistent adversarial networks}. In \bibinfo{booktitle}{{\em
  Proceedings of the IEEE international conference on computer vision}}.
  \bibinfo{pages}{2223--2232}.
\newblock


\end{thebibliography}

\end{document}